\documentclass[%
 aip,
 jmp,%
 amsmath,amssymb,
 reprint,%
]{revtex4-2}
\usepackage{graphicx,color,mhchem}
\graphicspath{{./}{v2_figures/}}
\usepackage{dcolumn}
\usepackage{bm}
\usepackage{placeins}
\usepackage[T1]{fontenc}
\usepackage[utf8]{inputenc}
\usepackage[margin=2.4cm]{geometry}
\usepackage{amsmath}
\usepackage{amssymb}
\usepackage{booktabs}
\usepackage{array}
\usepackage{siunitx}
\usepackage[colorlinks=true,linkcolor=black,citecolor=blue,urlcolor=blue]{hyperref}

\begin{document}

\title[black diamond]{Orientation-selective microwave heating of laser-graphitised %HPHT 
diamond at 2.45~GHz}

\author{Massimiliano Zamengo$^{1,*}$,
Meguya Ryu$^{1}$,
Haoran Mu$^{2,3}$, Hsin-Hui Huang$^{2,3}$, Nguyen Hoai An Le$^{2}$,
Tomas Katkus$^{2}$, Darius Gailevi\v{c}ius$^4$,
Junko Morikawa$^{1,5,6,*}$,
Saulius Juodkazis$^{2,4,5,*}$}

\affiliation{School of Materials and Chemical Technology,
Institute of Science Tokyo, 2-12-1, Ookayama, Meguro-ku, Tokyo 152-8550, Japan}
\affiliation{Optical Sciences Centre, School of Science, Swinburne University of Technology, Hawthorn, Victoria 3122, Australia}
\affiliation{Melbourne Center for Nanofabrication (MCN), 151 Wellington Road, Clayton, Vic 3168, Australia}
\affiliation{Laser Research Center, Physics Faculty, Vilnius University, Saul\.{e}tekio Ave. 10, 10223 Vilnius, Lithuania}
\affiliation{~World Research Hub (WRH), School of Materials and Chemical Technology, Institute of Science Tokyo, 2-12-1, Ookayama, Meguro-ku, Tokyo 152-8550, Japan}
\affiliation{~Research Center for Autonomous Systems Materialogy (ASMat), Institute of Innovative Research, Institute of Science Tokyo, Yokohama 226-8501, Japan}
\thanks{Correspondence: zamengo.m.82d4@m.isct.ac.jp (M.Z.);   morikawa.j.4f50@m.isct.ac.jp (J.M.); sjuodkazis@swin.edu.au (S.J.) }

\date{\today}

\begin{abstract}
Pristine diamond couples negligibly to a \SI{2.45}{GHz} field. We show that femtosecond-laser inscription of a buried graphitic (sp$^2$) architecture makes a $4\times4\times2$~mm$^3$ N-doped (\SI{100}{ppm}) high-pressure high-temperature (HPHT) crystal absorb strongly, and that the absorption is switched by the orientation of the inscribed structure relative to the cavity electric field. With the \SI{1.8}{mm} depth of a concentric grating along $\mathbf{E}$, stepped power gave a smooth staircase to an indicated $\sim 500^\circ$C at \SI{80}{W} followed, at \SI{90}{W}, by an abrupt excursion to an indicated $\sim 900^\circ$C with surface roughening and terraced oxidation. Rotating the same sample so that the grating diameter lay along $\mathbf{E}$, and an unmodified crystal in either orientation, gave no signal above the $\sim 40^\circ$C floor. We account for this by the depolarisation factors $L$ of the sub-wavelength tracks, collapsed into one screening parameter $S = L\sigma/(\omega\varepsilon_0\varepsilon_h)$, here, $\sigma$ conductivity, $\omega$ - cyclic frequency, $\varepsilon_{h,0}$ - permittivity of host and vacuum, respectively. The model %attributes the effect almost entirely to the single continuous axial line, predicts a conductivity optimum at $S=1$ near the resistivities actually achieved by laser writing, and 
turns element length, diameter and writing regime into design variables for microwave absorption in a transparent host.
\end{abstract}

\maketitle

\section{Introduction}
\label{s-intro}

A tightly focused ultrashort laser pulse drives a local sp$^3\!\rightarrow$\,sp$^2$ phase transition
inside diamond, and by translating the focus an electrically conductive graphitic wire of essentially arbitrary three-dimensional path can be written directly in the bulk~\cite{sotillo2016,kononenko2017}. The conductivity that results is governed less by peak
intensity than by how much heat accumulates around the focal volume, which makes the pulse-delivery format the controlling parameter. Single sub-picosecond pulses graphitise and quench a compact, near-diffraction-limited globule with little collateral damage but modest ordering~\cite{kuriakose2023}, whereas picosecond pulses and, more effectively, picosecond \emph{bursts} of femtosecond sub-pulses hold the excited volume above threshold long enough for fuller graphitisation. It is in this heat-accumulation regime that the lowest resistivities $\rho$ are reached: of order \SI{0.01}{\ohm\cdot cm} for the longest ($\sim$\SI{46.7}{ps}) sub-THz bursts, against $\rho \approx \SI{0.02}{\ohm\cdot cm}$ for single \SI{10}{ps} pulses and consistently higher values in the standard single-femtosecond-pulse regime~\cite{kuriakose2024,jedrkiewicz2017,momgaudis2025}. Reported extremes approach polycrystalline graphite --- \SI{0.022}{\ohm\cdot cm} in buried wires and \SI{0.75}{\ohm\cdot cm} in optimised multi-pass surface writing --- cutting the series resistance of bulk diamond by more than ten orders of magnitude~\cite{sotillo2016}. Raman and infrared spectroscopy read out the transformation: the single sharp sp$^3$ line at \SI{1332}{cm^{-1}} gives way to the graphitic G band near \SI{1580}{cm^{-1}} and the disorder-activated D band near \SI{1350}{cm^{-1}}, whose integrated intensity ratio $I_D/I_G$ tracks sp$^2$ cluster size and therefore correlates with the measured
conductivity~\cite{ferrari2000,sotillo2016}. The dark colour and finite conductivity of the resulting sp$^2$/sp$^3$ composite invite a loose comparison with natural carbonado~\cite{carbonado1998,blackdiamond2019}, though a laser-written single crystal is not the structural equivalent of a porous polycrystalline aggregate.

What has not been asked of these structures is how they behave in a microwave field. A dielectric in an oscillating field dissipates a time-averaged power density $p$ (in \si{W/m^3}) given by~\cite{bykov2001jpd,Horikoshi2024}
\begin{equation}
p = 2\pi f\,\varepsilon_0\,\varepsilon''_{\mathrm{eff}}\,|E|^2, \qquad
\varepsilon''_{\mathrm{eff}} = \varepsilon''_{\mathrm{dip}} + \frac{\sigma}{2\pi f \varepsilon_0},
\label{e-power}
\end{equation}
where $f$ is the frequency (here \SI{2.45}{GHz}), $\varepsilon_0$ the permittivity of free space, $|E|$ the root-mean-square local field amplitude, $\varepsilon''_{\mathrm{eff}}$ the effective
relative loss factor, $\varepsilon''_{\mathrm{dip}}$ its dipolar-polarisation part and $\sigma$ the
electrical conductivity (in \si{S/m}), with $\rho = 1/\sigma$ the corresponding resistivity. The two terms are the two ways a solid can absorb: reorientation of permanent or defect-associated dipoles,
and Ohmic dissipation by mobile charge. Both are exceptionally weak in diamond. The lattice is non-polar, so $\varepsilon''_{\mathrm{dip}}$ is small and the loss tangent
$\tan\delta = \varepsilon''_{\mathrm{eff}}/\varepsilon_h$ with $\varepsilon_h = 5.7$ the relative
permittivity of the host --- is generally quoted in the $10^{-4}$--$10^{-6}$ range at
\SI{2.45}{GHz}. Free carriers are scarce because the dominant impurity in type-Ib material, single substitutional nitrogen, is a \emph{deep} donor lying $E_d = \SI{1.7}{eV}$ below the conduction
band~\cite{farrer1969}: even at $N_s \approx \SI{1.8e19}{cm^{-3}}$ ($\sim$\SI{100}{ppm}) the
room-temperature free-electron density is $n \sim \SI{e5}{cm^{-3}}$ and the crystal is
semi-insulating. Its thermal conductivity $k$ is moreover the highest of any bulk solid, so whatever power is deposited is rapidly spread. On every count a diamond block in a \SI{2.45}{GHz} cavity should
not heat.

Laser inscription supplies, at chosen positions and with chosen orientation, exactly the ingredient that is missing: mobile charge. This places the present experiment in a long line of work in which
microwave absorption is engineered rather than inherited~\cite{Osep,sutton1989,bykov2001jpd,Horikoshi2024,Wang2026}. Since the foundational demonstrations of alumina sintered at \SI{28}{GHz} some \SIrange{250}{500}{\celsius} below the conventional schedule~\cite{janney1988,janney1990,janney1991,brosnan2003}, and full densification of powdered-metal bodies once the particle size approached the skin depth $\delta_s$~\cite{roy1999}, the recurring lesson has been that coupling is set by microstructure and geometry rather than by chemistry alone, and that it evolves as the sample heats. Where the
intrinsic loss is too small, susceptors and hybrid schemes supply it~\cite{bhattacharya2016,carbon2014}; where it is too large, runaway and hot-spotting follow.
Semiconductors sit in between and traverse the range during an anneal, which is the basis of the
low-thermal-budget microwave activation now demonstrated for silicon, germanium, SiC and
beyond~\cite{sic2007,lowtemp2013,ge_apl2016,geonsi2016,nanosheet2022,Truong2025}. %and of 
The continuing discussion proceeds whether the field contributes anything beyond heat, a ponderomotive drift
on hopping defects that becomes appreciable only where the gradient $\nabla|E|^2$ is large~\cite{rybakov1994,rybakov1995,rybakov1997,booske1997,srep2016,nonthermal_review2025}. The practical obstacle throughout has been not the physics but spatial control of where the energy lands~\cite{susceptor2012,mechanistic2025}.

Laser writing bears directly on that obstacle, and diamond is an unusually clean testbed: a low-loss
host means the microwave response can be attributed to the inscribed sp$^2$ network, whose position,
length, spacing and orientation are set by the writing beam. Microwave energy has so far entered
diamond science from a different direction: microwave-plasma CVD grows the crystals, while lattice
damage and vacancy mobilisation are repaired by high-temperature vacuum
annealing~\cite{nvimplant2017}. Whether a volumetric, field-addressed alternative is even
possible has not been tested. This paper establishes that it is, and identifies what controls it. 

We show that a graphitised concentric grating written inside a $4\times4\times2$~mm$^3$ N-doped high-pressure high-temperature (HPHT) block allows the crystal to be driven, at a launched power $P_{\mathrm{in}} \sim \SI{90}{W}$ in a \SI{2.45}{GHz} cavity, into strong absorption with surface roughening and terraced oxidation, and that this happens only when the $\sim$\SI{1.8}{mm} depth of the structure is aligned with the microwave electric field $\mathbf{E}$: %rotating 
positioning the sample so that the grating diameter lies along $\mathbf{E}$, or testing an unmodified crystal, gives no detectable response.
Section~\ref{s-results} reports the observations and Sec.~\ref{s-discussion} shows that they follow
from the depolarisation factors of the sub-wavelength tracks, which reduce the orientation dependence
to a single screening parameter and single out the central axial line as the active element.
Inscription parameters, the buried graphite/diamond contact and the calibration of the thermometry
are collected in Appendices~\ref{a-writing}--\ref{a-emis}.

%____________________________fig 1
\begin{figure*}[t]
\centering\includegraphics[width=0.85\textwidth]{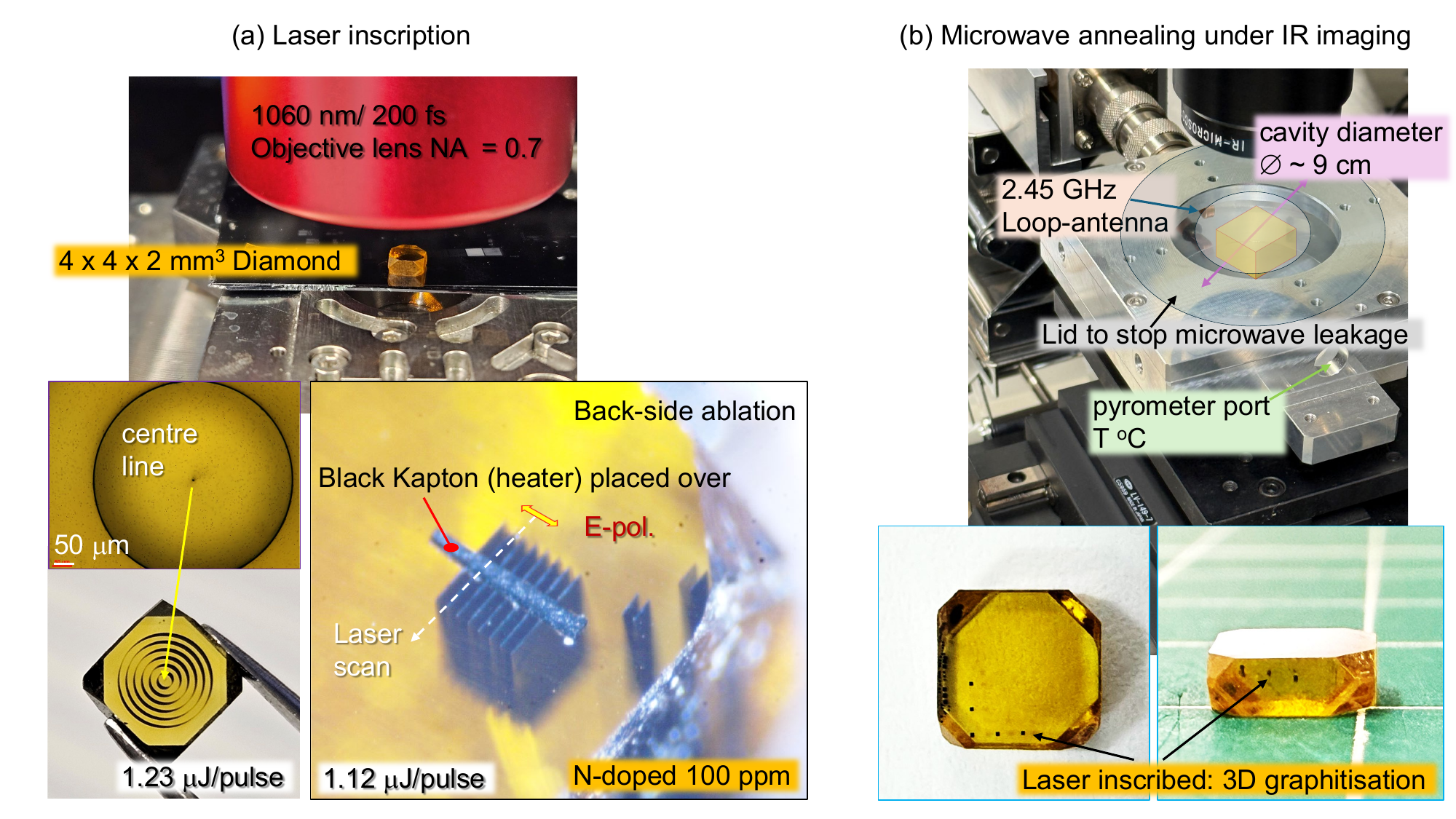}
\caption{\label{f-setups} (a) Fs-laser (\SI{1030}{nm}/\SI{200}{fs}) inscription starting from the
back-side face, with optical images of the N-doped HPHT diamond and its inscribed graphitised
regions. Writing parameters are collected in Table~\ref{t-writing} of Appendix~\ref{a-writing}.
(b) Microwave cavity oven (low-depth version) used for imaging the IR emissivity of the
laser-structured regions; the loop antenna injects the microwave field into the sample chamber. The
$E$-field is perpendicular to the top plane of the chamber, i.e.\ on the axis of the cylindrical
chamber. The 50-$\mu$m-thick black Kapton tape slice (a) and laser-inscribed $0.1\times 0.1\times 0.1$~mm$^3$ micro-volumes (b) are for a laser-driven heater function of the thermal wave experiments~\cite{RYU2020120337} (not reported here). }
\end{figure*}

\section{Experimental}
\label{s-exp}

\subsection{Host crystal and laser inscription}

\begin{figure*}[t]
\centering\includegraphics[width=0.85\textwidth]{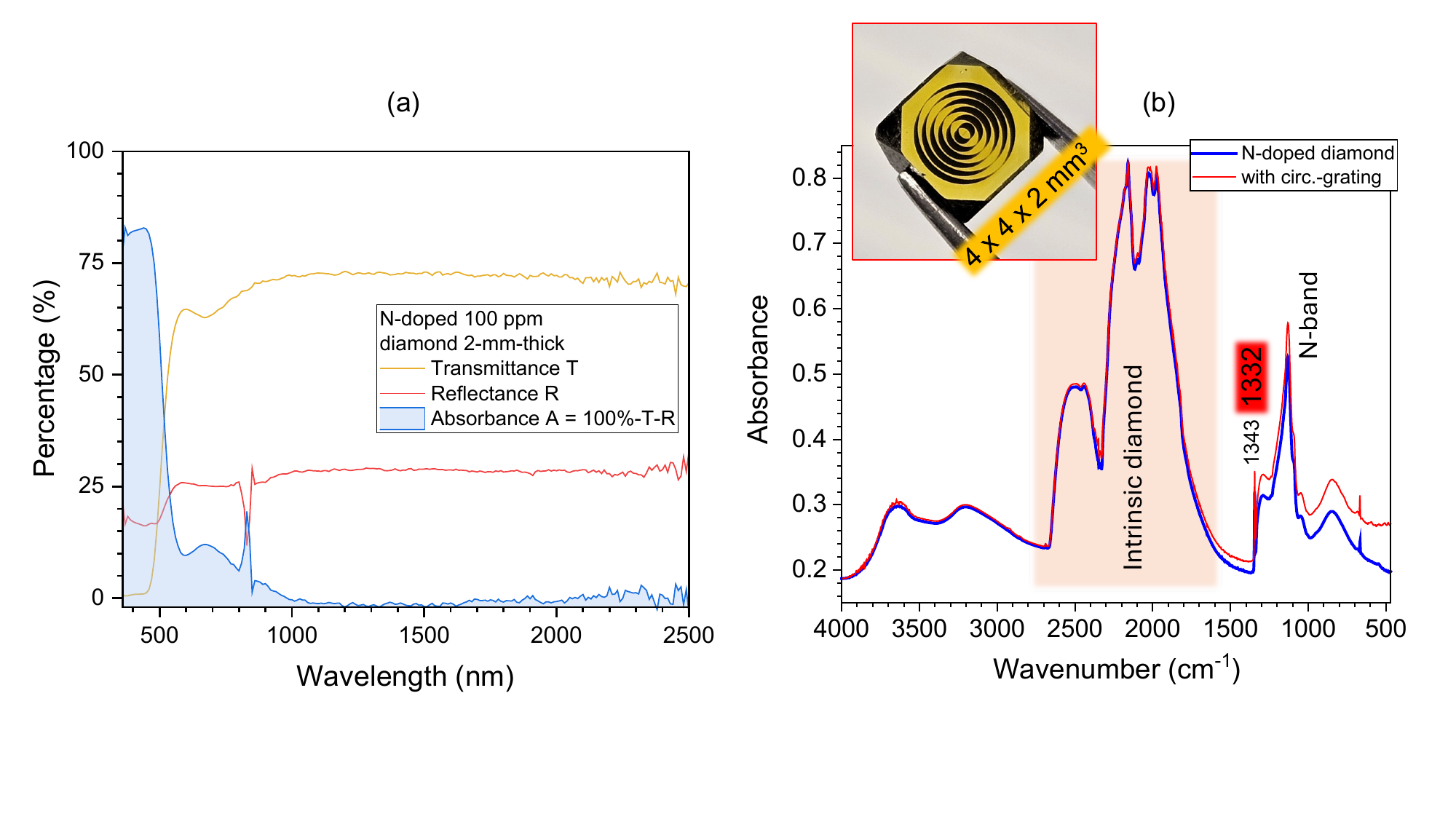}
\caption{\label{f-abs} (a) Transmittance $T$, reflectance $R$ and absorptance $A = 1-R-T$ of a
laser-untreated HPHT \SI{100}{ppm} N-doped diamond, $4\times 4\times 2$~mm$^3$; $T$ and $R$ were measured directly. (b) IR spectra of pristine and laser-inscribed (inset photo) samples. The one-phonon feature near \SI{1343}{cm^{-1}} is attributed to the N$_s^0$ local mode of high-concentration substitutional nitrogen.  Spectrometer calibration is given in Appendix~\ref{a-calib}, Fig.~\ref{f-cali}.}
\end{figure*}

The host is a synthetic type-Ib HPHT crystal of $4\times4\times2$~mm$^3$ containing
$\sim$\SI{100}{ppm} single substitutional nitrogen (Infi Advanced Materials Co.,Ltd.), whose absorption and luminescence are dominated by the neutral C-centre (N$_s^0$): absorption ramps up from a threshold near \SI{2.2}{eV} ($\sim$\SI{560}{nm}) toward the ultraviolet together with the characteristic \SI{270}{nm} band, giving the crystal its yellow colour~\cite{green2022rimg,jones2009,huang2022ns}. In the one-phonon infrared region the same defect produces the N$_s^0$ features near \SI{1130}{cm^{-1}} and \SI{1344}{cm^{-1}}, with \SI{1332}{cm^{-1}} marking the ionised donor N$_s^+$~\cite{green2022rimg}; the \SI{1344}{cm^{-1}} feature is a nitrogen-related local mode and not the intrinsic \SI{1332}{cm^{-1}} LO/TO phonon displaced by doping, a distinction that matters for Figs.~\ref{f-abs} and \ref{f-ftir}. Because HPHT growth uses a nickel-based solvent/catalyst, charge
transfer (N$_s^0 + \mathrm{Ni} \rightarrow \mathrm{N}_s^+ + \mathrm{Ni}^-$) couples the two impurity systems and populates the near-infrared with nickel- and nickel--nitrogen-related centres, alongside
NV and N--V--N features wherever vacancies are
present~\cite{nazare1991,yelisseyev2007,zaitsev2001,green2022rimg,davies1976nv,davies2013nv0,%
goss1996nv}. The relevant consequence here is that nickel \emph{compensates}: it converts N$_s^0$
into N$_s^+$ and so reduces the neutral-donor reservoir available for the thermal ionisation
considered in Sec.~\ref{s-nitrogen}, and its concentration was not determined. Transmittance,
reflectance and inferred absorptance of the untreated crystal --- denoted $T$, $R$ and $A = 1-R-T$
in the caption of Fig.~\ref{f-abs} only, where $T$ is not the temperature together with infrared
spectra of pristine and laser-inscribed material are shown in Fig.~\ref{f-abs}.

Figure~\ref{f-setups}(a) shows the inscription geometry; Table~\ref{t-writing} of Appendix~\ref{a-writing} collects the writing parameters. An objective of numerical aperture $\mathrm{NA} = 0.7$ focused \SI{1030}{nm} pulses of duration $t_p = \SI{200}{fs}$, starting at the back face of the transparent sample and stepping toward the front. The structure consists of a set of concentric rings written by scanning, together with a central axial line written with the beam stationary at 2000 pulses per pass. Forty passes at an axial step $\Delta z = \SI{50}{\micro\meter}$ produced a structure $\sim$\SI{1.8}{mm} deep, the last seven passes at linearly reduced power ($100\% - 11.9\%\times N_{\mathrm{p}}$, $N_{\mathrm{p}} = 1\ldots7$) to avoid cracking. Both the
rings and the central column are continuous: the axial extent of the volume modified under these
conditions substantially exceeds the nominal depth of focus due to combined spherical aberration and self-focusing. So successive passes merge and the rings close into loops rather than dashed arcs as determined by optical microscopy. %This continuity is the structural fact on which the analysis of Sec.~\ref{s-discussion} rests most heavily, and its direct verification is the highest-value measurement still outstanding (Sec.~\ref{s-scope}). 
Visible marks appeared on the
back side but not on the front, which lay $\sim$\SI{144}{\micro\meter} below the end of the
structure.

Diffraction-limited focusing would give a focal spot diameter
$2r = 1.22\lambda_L/\mathrm{NA} = \SI{1.8}{\micro\meter}$ and a depth of focus $2z_R = \SI{4.9}{\micro\meter}$, with the nominal fluence and irradiance listed in Table~\ref{t-writing}; these are upper bounds only, since focusing through millimetres of $n \approx 2.4$ material with a dry objective introduces uncorrected spherical aberration that elongates the interaction volume axially, the same effect that makes the observed axial continuity plausible despite the \SI{50}{\micro\meter} step. The quantity that enters the microwave analysis is therefore the \emph{observed} track diameter $a = \SIrange{5}{10}{\micro\meter}$, measured directly.
The buried graphite/diamond boundary is a metal--semimetal contact discussed in Appendix~\ref{a-junction}; none of the analysis below depends on it, because an isolated conductive inclusion dissipates energy through its own internal conduction current and needs no ohmic connection to the host.

%____________________________fig 2

\subsection{Microwave cavity and diagnostics}

Microwave heating and dielectric characterisation used a cylindrical TM$_{010}$ resonator operating from 2.4 to \SI{2.5}{GHz} (Minamo, Japan), connected through a waveguide, slug tuner and coaxial cables to a signal generator (MR-2G-100, Ryowa Denshi, Japan); impedance matching and frequency tracking were handled in software so that resonance was maintained as the load changed. The setup~\cite{Massi} launches up to $P_{\mathrm{in}} = \SI{100}{W}$ at $f = \SI{2.45}{GHz}$, free-space wavelength $\lambda_0 = \SI{12.24}{cm}$, with a tunability window of $\pm$\SI{50}{MHz} about the centre frequency. Resonant frequency and quality factor $Q$ were obtained by pseudo-Voigt fitting of vector network analyser data (VNA; SV4401A, Sysjoint, China); $Q$ was not logged synchronously with the heating run, which is why the analysis below uses frequency only. For an ideal empty cylindrical TM$_{010}$ cavity, $f = x_{01}c/(2\pi R)$ with $x_{01} \approx 2.405$ gives a diameter $2R \approx \SI{9.37}{cm}$ at \SI{2.45}{GHz}, i.e.\ $0.765\lambda_0$ rather than $\lambda_0$.

Temperature was monitored with a radiation thermometer (TMHX-STM0050-0070E003, Japan Sensor, Japan) reading the sample through a \ce{SiO2} tube over the 1.95--2.6~$\mu$m band with an InSb sensor, calibrated against K-type thermocouples up to $250^\circ$C (Appendix~\ref{a-calib}, Fig.~\ref{f-tk}). The sensing spot was \SI{3}{mm} in diameter at \SI{7}{cm} depth into the cavity. The instrument emissivity setting and the tube transmission correction were held fixed throughout the run, so all temperatures reported below are \emph{indicated} radiative temperatures;
Appendix~\ref{a-emis} quantifies how far these may depart from true surface temperature and is referred to wherever a temperature is interpreted. The \ce{SiO2} tube was open to room air without forced flow. The sample was supported and oriented relative to $\mathbf{E}$ using quartz \ce{SiO2} wool (2--4~$\mu$m fibres), which did not measurably perturb the cavity resonance.

A finite-element model of the loaded cavity (COMSOL Multiphysics 6.4) was used for qualitative comparison of the field and loss topology; it is described in Sec.~\ref{s-comsol}.

\section{Results}
\label{s-results}

\subsection{Orientation-selective heating and an abrupt transition}
\label{s-orient}

\begin{figure*}[t]
\centering\includegraphics[width=0.72\textwidth]{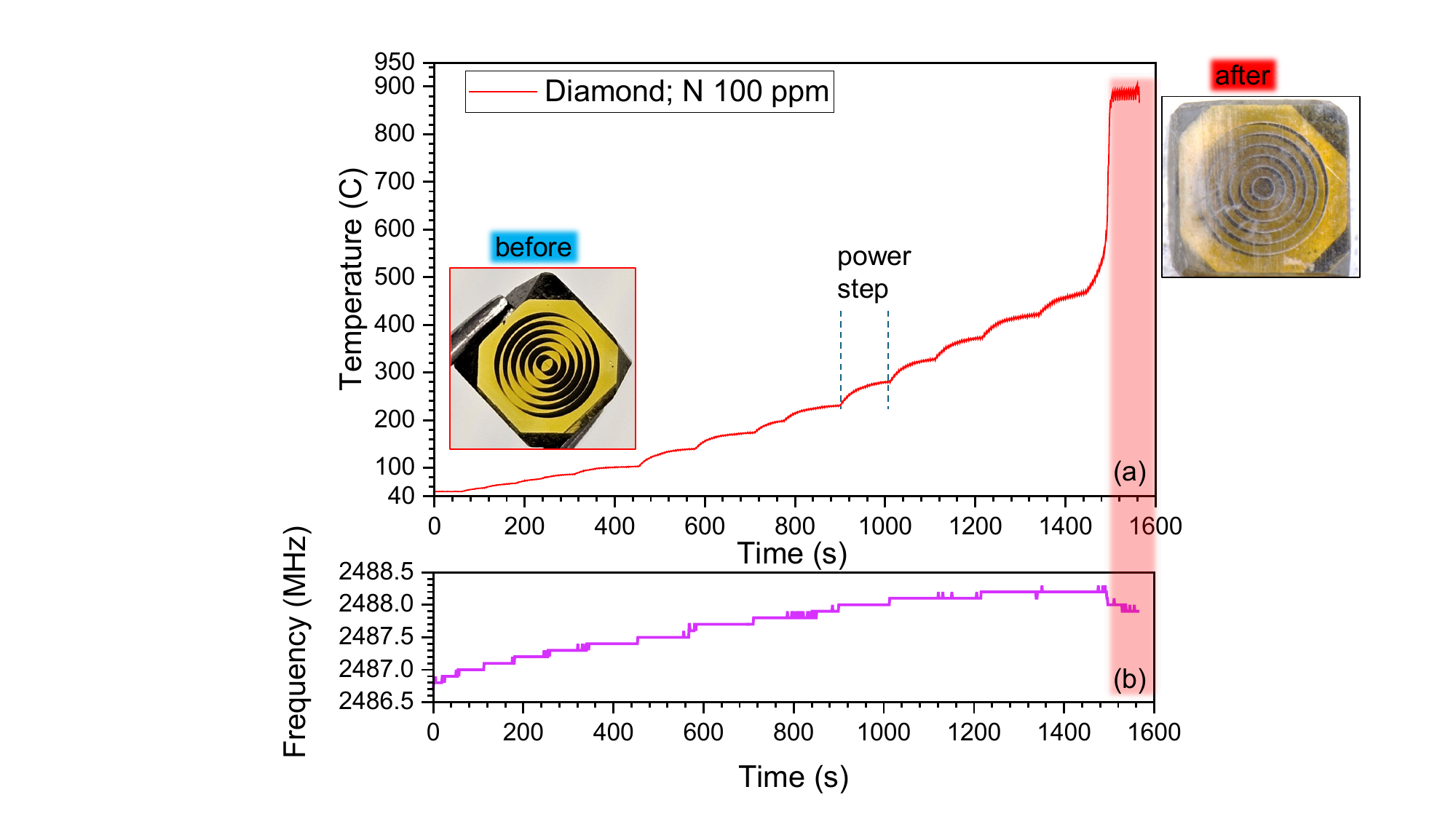}
\caption{\label{f-hta} (a) Pyrometer-measured radiative temperature on the sample top surface inside
the quartz tube. (b) The corresponding resonant-frequency change during annealing. Power steps were
5--\SI{10}{W}, with the abrupt transition at \SI{90}{W}. When the microwave $E$-field was
perpendicular to the depth axis of the inscribed cylindrical structure, no heating was detected
(below the $40^\circ$C floor of measurable temperature rise). Insets show photographs of the same
sample before the experiment and after surface oxidation. Indicated temperatures should be read
together with Appendix~\ref{a-emis}.}
\end{figure*}

Figure~\ref{f-hta}(a) shows the indicated radiative temperature of the diamond as the launched power
$P_{\mathrm{in}}$ was ramped in 5--\SI{10}{W} steps from \SI{5}{W} to \SI{90}{W}, with the concentric grating aligned with the cavity electric field $\mathbf{E}$. Each power step produces a distinct plateau and the trajectory is smooth up to an indicated $\sim 500^\circ$C at $P_{\mathrm{in}} = \SI{80}{W}$. The next \SI{10}{W} increment triggers an abrupt excursion to an indicated $\sim 900^\circ$C, reached within tens of seconds rather than over the several-minute settling time of the preceding steps. Heating was terminated by the operator at that point because
reflected power was rising. The photographic insets document the accompanying change from a clear yellow crystal to a visibly roughened one.

The control measurements define the effect. An unmodified diamond block, and the same inscribed sample rotated so that the grating diameter rather than its $\sim$\SI{1.8}{mm} depth lies along $\mathbf{E}$, both remained below the $\sim 40^\circ$C floor of the measurement at all powers up to
$P_{\mathrm{in}} = \SI{90}{W}$. The response is therefore not a property of diamond, nor of
laser-modified diamond alone, but of laser-modified diamond \emph{in a particular orientation}. This is the central observation of the paper, and Sec.~\ref{s-discussion} shows that the orientation is precisely the variable that a depolarisation argument predicts should control the coupling.

Two qualifications attach to the measurement itself. These are single runs on a single principal specimen, and the heating run is destructive and hence not repeatable on the same sample. Rotating the sample also changes its position within the axial field, its contact with the quartz-wool support, the pyrometer viewing geometry and the cavity loading, so the contrast reported here is a comparison of two configurations rather than a controlled angular scan, which would require a precision fixture holding the sample centre fixed and a continuous sweep of angle.

%____________________________fig 3

\subsection{Cavity resonance during the run}
\label{s-freq}

Figure~\ref{f-hta}(b) shows the cavity resonance tracked simultaneously. The resonant frequency rises
monotonically through the stepped ramp, by $\Delta f \approx \SI{1.7}{MHz}$ in total, mirroring the
temperature staircase step for step, and then reverses at the onset of the abrupt excursion. The
reversal coincides with the rise in reflected power. Frequency alone carries mainly reactive
information --- it responds to the sample permittivity and conductivity, but equally to thermal expansion of the cavity, displacement of the sample or its support, changes in coupling through the tuner, and oxidation-induced changes of sample geometry --- while loss information resides in $Q$, which was not recorded synchronously. The step-for-step correspondence with the temperature staircase is nevertheless the strongest available indication that the load itself, and not the instrumentation, is what changes through the ramp, and the reversal marks a qualitative change in the character of that load at the transition.

\subsection{Post-run morphology and surface oxidation}
\label{s-morph}

\begin{figure*}[t]
\centering\includegraphics[width=0.70\textwidth]{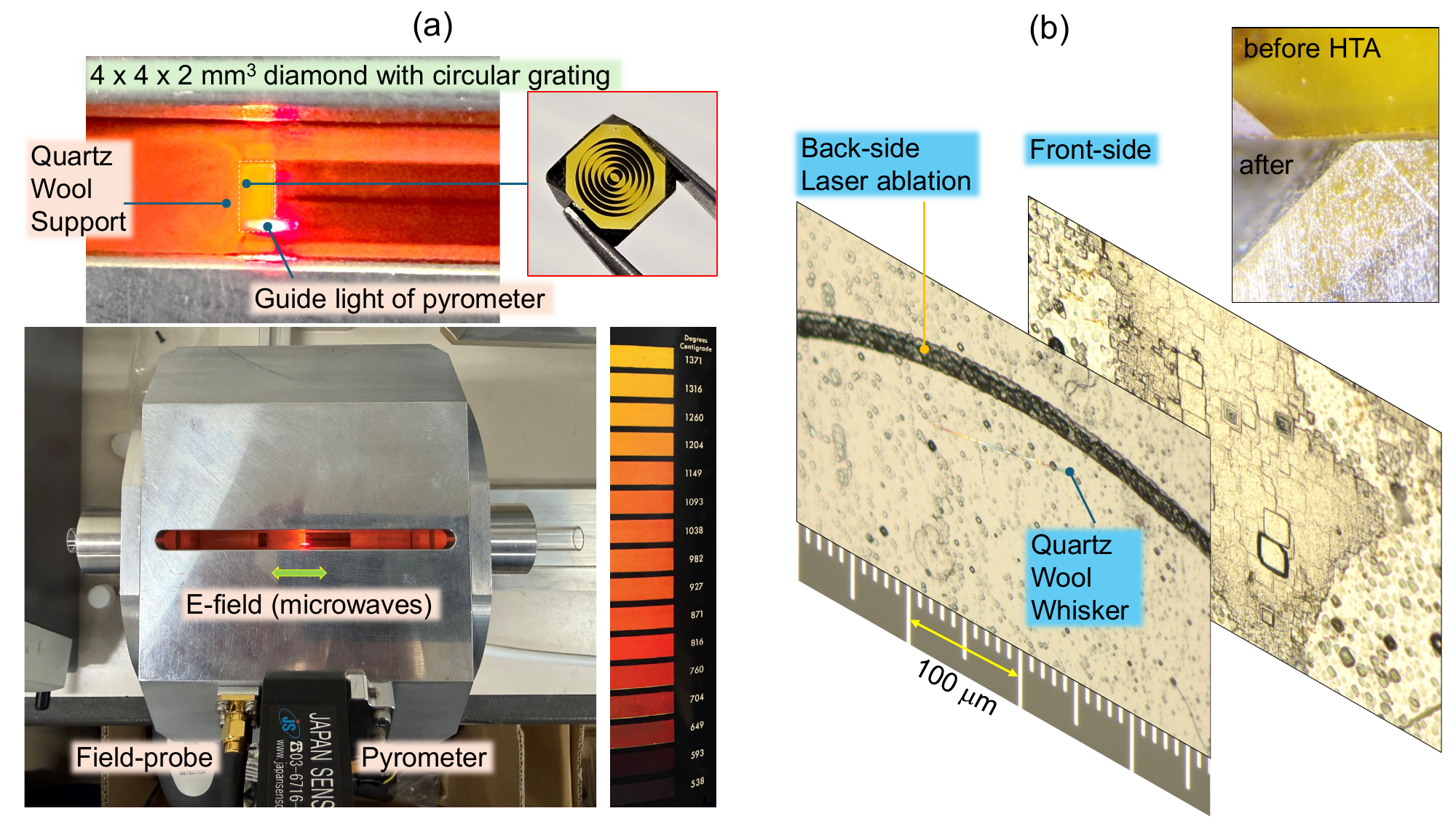}
\caption{\label{f-MW} (a) Annealing of the diamond with the inscribed circular grating inside the microwave oven (large-depth version); the microwave field is injected into the sample cavity via the waveguide. Direct optical observation at the conditions at the end of the heating experiment of Fig.~\ref{f-hta}; the radiative temperature scale is shown at right~\cite{cbar}. (b) Microscopy images of the front and back surfaces of the sample after annealing. The photo-inset shows the change produced by annealing and surface oxidation.}
\end{figure*}

Figure~\ref{f-MW}(a) shows direct real-colour observation of the sample at the condition at which the run was stopped, incandescent along its length in the cavity, together with the radiative temperature scale. Figure~\ref{f-MW}(b) shows the front and back surfaces afterwards. A clear pattern of terraces is recognisable, consistent with thermally driven oxidation proceeding preferentially along crystallographic planes and from defect sites, and the photographic inset shows the corresponding change in appearance of the whole crystal. No mass loss, surface recession or exhaust gas analysis was performed, so the extent and products of the oxidation are not quantified. Two consequences carry into the interpretation: the surface that the pyrometer views is not the surface it viewed at the start of the run (Appendix~\ref{a-emis}), and carbon oxidation is exothermic, so chemical heat release is a contributor to the transition that these data do not separate from dielectric heating (Sec.~\ref{s-scope}).

%____________________________fig 4

\subsection{Where the graphitic material lies}
\label{s-spectro}

\begin{figure*}[t]
\centering\includegraphics[width=0.9\textwidth]{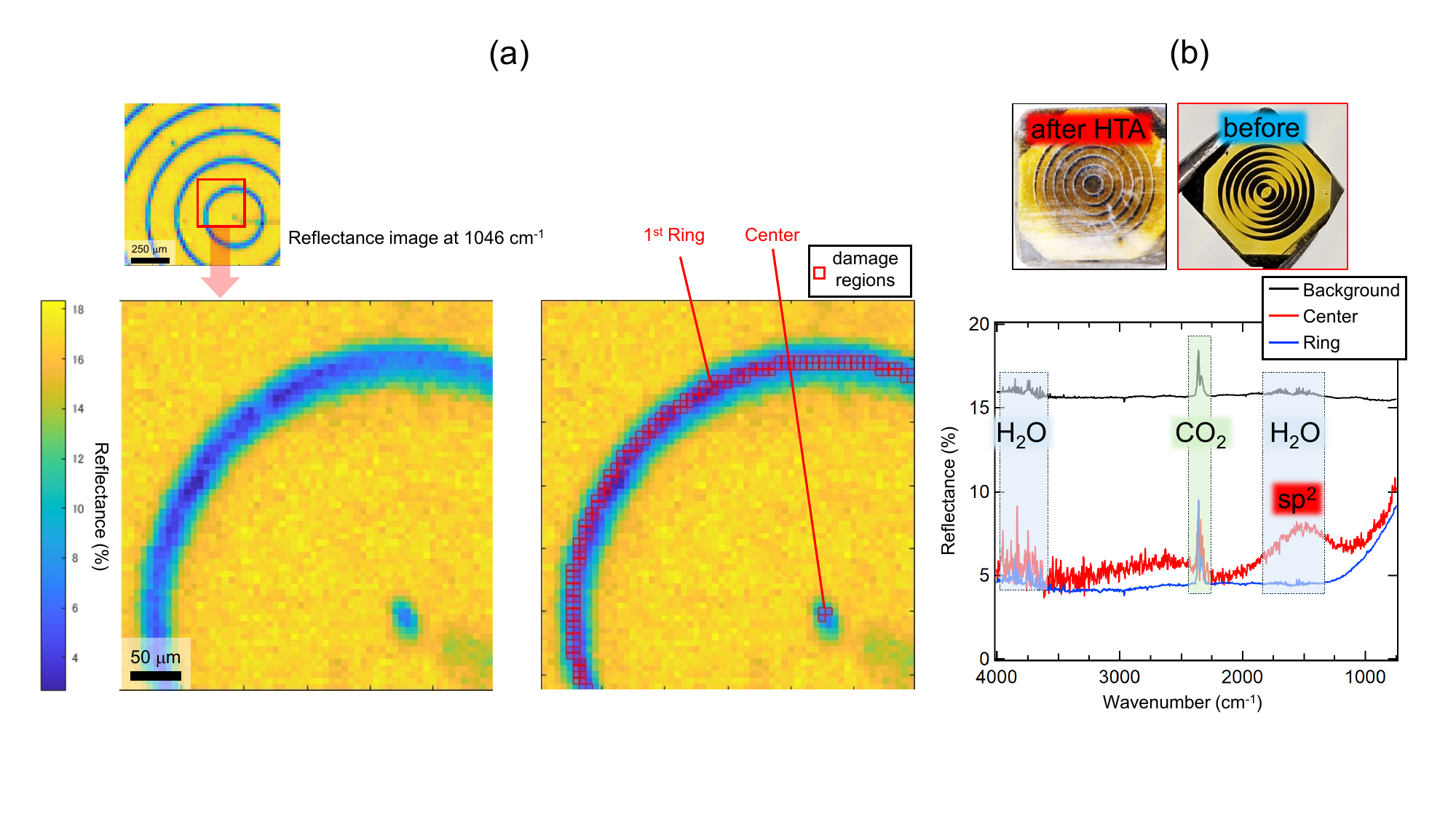}
\caption{\label{f-ftir} IR image and spectra measured at $10~\mu$m resolution (objective lens of numerical aperture $\mathrm{NA} = 0.6$). Averaging of 2--3 pixels showed the presence of the graphitic \SI{1500}{cm^{-1}} band on the centre line, and also at the level of single pixels along the scanned circles.}
\end{figure*}

Infrared reflectance was mapped across the laser-inscribed regions at $10~\mu$m resolution (Fig.~\ref{f-ftir}). Three features are reproducibly present in the modified regions. The broad band near \SI{1550}{cm^{-1}} is a G-like infrared-active sp$^2$ mode, related to the infrared-active graphitic mode near \SI{1590}{cm^{-1}}; it is evidence for graphitic bonding, but not by itself a measure of disorder, and it is not the Raman D band, which lies near \SI{1350}{cm^{-1}}~\cite{ferrari2000}. The feature near \SI{1110}{cm^{-1}} cannot be assigned to a vacancy-related sp$^3$ defect here, because it overlaps the broad one-phonon N$_s^0$ absorption near \SI{1130}{cm^{-1}} intrinsic to this heavily doped host~\cite{green2022rimg}. The feature near
\SI{1720}{cm^{-1}} is consistent with a carbonyl \ce{C=O} stretch and hence with an oxygen-containing surface species, though not with \ce{CO2} production, for which no gas-phase measurement was made. The measurement was unpurged and the measured quantity is reflectance, which maps onto absorption only through the complex refractive index, so no single-pixel defect assignment is attempted.

The spatial information is the most useful content of the figure and it is directly relevant to the mechanism. The sp$^2$-related reflectance is modulated on a scale requiring the $10~\mu$m resolution to resolve, and its strongest expression is on the central line, written with the beam stationary; the rings, written by scanning, show the same signature more weakly. The central line is thus both the most strongly graphitised element and, as Sec.~\ref{s-geometry} shows, the element that the depolarisation model identifies as carrying essentially all of the orientation-selective absorption.

%____________________________fig 5

\section{Discussion}
\label{s-discussion}

\subsection{Sub-wavelength conductive inclusions and the screening parameter}
\label{s-S}

At $f = \SI{2.45}{GHz}$ the free-space wavelength is $\lambda_0 = \SI{12.24}{cm}$ and, inside diamond, $\lambda = \lambda_0/\sqrt{\varepsilon_h} = \SI{5.13}{cm}$. The inscribed elements, at most \SI{2}{mm} in any dimension, are therefore $\lambda/25$ or smaller: they are not resonant structures and their response is quasi-electrostatic. For an inclusion of prolate or oblate spheroidal shape embedded in a host of relative permittivity $\varepsilon_h$, the internal field is uniform and given by~\cite{bohren1983,landau1984ecm,sihvola1999}
\begin{equation}
E_{\mathrm{in}} = \frac{\varepsilon_h}{\varepsilon_h + L(\varepsilon_i - \varepsilon_h)}\,E_0,
\qquad \varepsilon_i \simeq \varepsilon_h - i\,\frac{\sigma}{\omega\varepsilon_0},
\label{e-depol}
\end{equation}
where $E_0$ is the field applied to the inclusion by the surrounding host, $\varepsilon_i$ the complex relative permittivity of the graphitic inclusion, $\omega = 2\pi f$, and $L$ the depolarisation factor along the direction of $E_0$. For a spheroid of symmetry-axis length $h$ and equatorial diameter $D$ the factors along and across the symmetry axis follow from the standard expressions~\cite{osborn1945,stoner1945}, reducing for a slender prolate body of aspect ratio
$m = h/D \gg 1$ to
\begin{equation}
L_\parallel \simeq \frac{\ln (2m) - 1}{m^{2}}, \qquad
L_\perp = \frac{1 - L_\parallel}{2},
\label{e-L}
\end{equation}
with the sum rule $L_\parallel + 2L_\perp = 1$ holding for all shapes.

Because the inclusion is conductive, $\varepsilon_i - \varepsilon_h = -i\sigma/(\omega\varepsilon_0)$
is purely imaginary and the denominator of Eq.~(\ref{e-depol}) collapses to
$\varepsilon_h - iL\sigma/(\omega\varepsilon_0)$. The whole orientation dependence is then carried by
a single dimensionless \emph{screening parameter},
\begin{equation}
S \equiv \frac{L\,\sigma}{\omega\varepsilon_0\varepsilon_h},
\qquad
\left|\frac{E_{\mathrm{in}}}{E_0}\right| = \frac{1}{\sqrt{1+S^{2}}}\,,
\qquad
p = \frac{\sigma|E_0|^{2}}{1+S^{2}},
\label{e-S}
\end{equation}
the last expression following from Eq.~(\ref{e-power}) with
$\varepsilon''_{\mathrm{eff}} = \sigma/(2\pi f\varepsilon_0)$. For $S \ll 1$ the inclusion is
effectively invisible to the field and $E_{\mathrm{in}} \to E_0$; for $S \gg 1$ the field is largely excluded and $E_{\mathrm{in}} \to E_0/S$; $S \approx 1$ is the crossover.

Physically, $S$ balances how much surface charge the applied field drives to the inclusion boundary against how effectively that charge, once there, builds a depolarising field opposing $E_0$ inside.
The second factor is purely geometric and is what $L$ measures. In a slender body driven along its long axis the induced surface charge sits at the two distant terminations, and the field such a widely separated distribution produces in the interior is weak. Driven across the short axis, the same charge sits on closely spaced opposite faces and the cancellation is nearly complete. This is the same geometry, and the same algebra, that makes a slender ferromagnet easy to magnetise along its length and hard across it~\cite{osborn1945,stoner1945}. An equivalent reading is $S = 1/(\omega\tau)$
with $\tau = \varepsilon_0\varepsilon_h/(L\sigma)$ the relaxation time for establishing the screening charge, to be compared with $\omega^{-1} = \SI{65}{ps}$ at \SI{2.45}{GHz}. It is worth noting that
$\tau$ is not a carrier transit time: surface-charge formation is a collective electrodynamic response, and $L$ encodes the geometry of the resulting charge distribution, not a distance travelled by individual carriers.

Two things that $S$ does not describe should be separated from it. The skin depth in the graphitic
material, $\delta_s = \sqrt{2/(\omega\mu_0\sigma)} \approx \SI{144}{\micro\meter}$ at
$\sigma = \SI{5000}{S/m}$, greatly exceeds the track diameter $a \approx \SI{7.5}{\micro\meter}$, so
there is no skin-effect variation of the field \emph{across} the conductor cross-section; the
depolarisation screening of Eq.~(\ref{e-S}) instead reduces the magnitude of the spatially uniform
internal field relative to the applied one, and a transversely driven track can have no skin-effect
gradient and a strongly suppressed internal field at the same time. And because the ellipsoid is the
unique shape for which the internal field is uniform, Eq.~(\ref{e-depol}) describes the
volume-integrated coupling and its orientation dependence, not the axial profile: a real element with
flat terminations must carry $J \to 0$ there by charge conservation, so its current density peaks
toward the centre, as the finite-element solution of Sec.~\ref{s-comsol} shows.

\subsection{Field and loss topology in the loaded cavity}
\label{s-comsol}

\begin{figure*}[t]
\centering\includegraphics[width=0.90\textwidth]{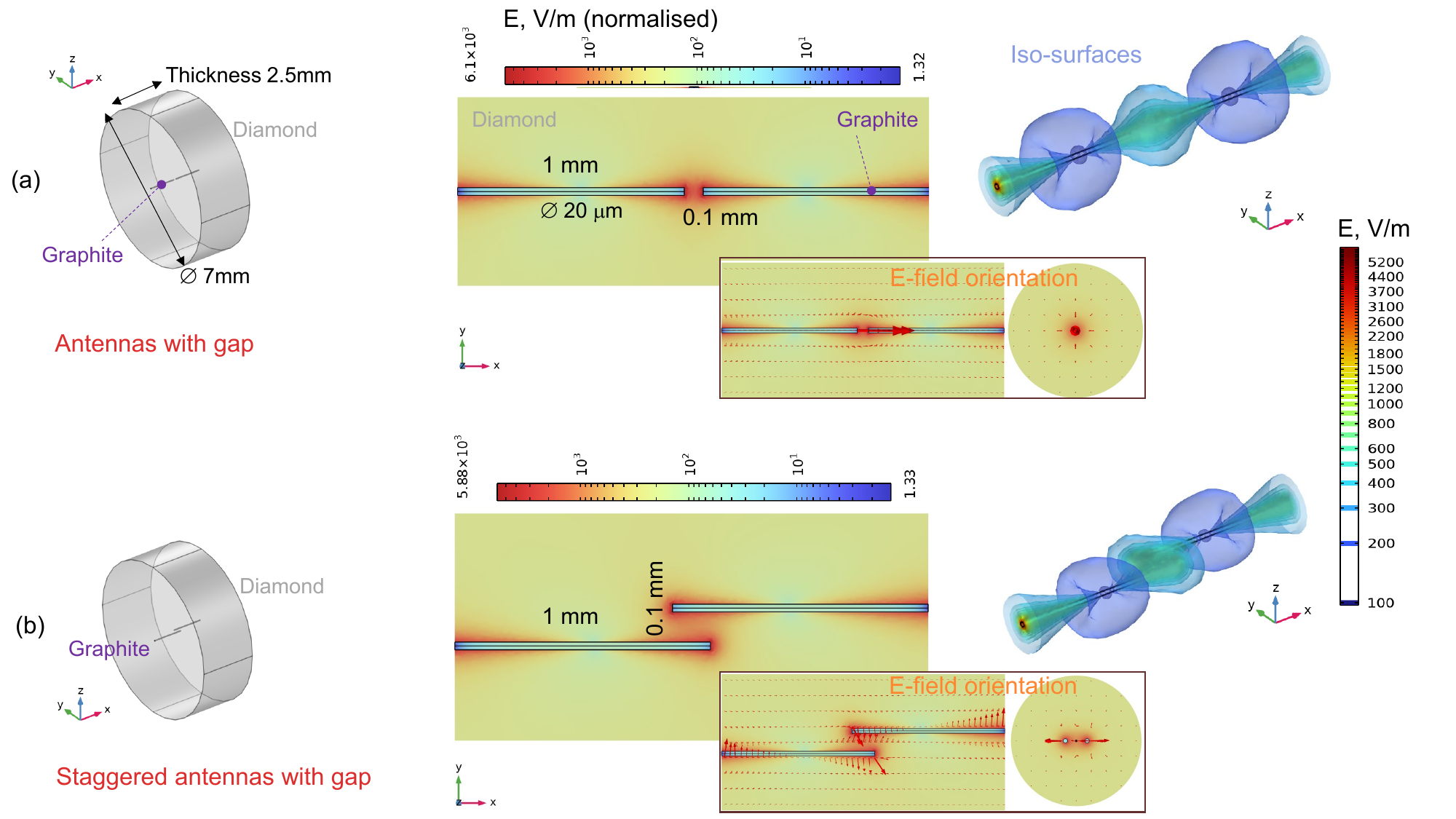}
\caption{\label{f-simu}  COMSOL Multiphysics 6.4 simulation of the microwave electric field in a diamond disk (diameter 7~mm, thickness 2.5 mm) containing laser-inscribed graphite wires (diameter 20~$\mu$m, length 1 mm each), placed on the axis of the TM010 cylindrical cavity. (a) Two collinear wires separated by an axial gap of 0.1 mm; (b) two parallel wires staggered by a lateral offset of 0.1 mm. For each configuration, from left to right: 3D sketch of the sample; cross-section of $|E|$ in the xz-plane through the wires (normalized to the maximum field); magnified view of the gap region with arrows indicating the E-field direction, and cross-section perpendicular to the wires (yz-plane); iso-surfaces of $|E|$ in the sample. 
Material properties: diamond, $\varepsilon_r = 5.5 - 0.001i$, $\sigma=1\times10^{-12}$~{S/m}; graphite, $\varepsilon_r = 10$,  $\sigma$ = 1000 S/m. The computed eigen frequencies are 2483.2 + 
0.418$i$ MHz (a) and 2483.1 + 0.418$i$ MHz (b).
}
\end{figure*}

Figure~\ref{f-simu} shows a finite-element solution (COMSOL Multiphysics 6.4) for the loaded cavity with a simplified inclusion geometry: a pair of collinear graphitic elements separated by a \SI{0.1}{mm} gap rather than the full three-dimensional concentric-ring structure. Assumed properties were relative permittivity $5.5-0.001i$ and conductivity \SI{1e-12}{S/m} for diamond, and relative permittivity 10 and conductivity \SI{1000}{S/m} for graphite, with an inscription diameter of $20~\mu$m. The electric field is largest at the element terminations and across the gap, while the loss density is largest toward the element centres --- the non-uniform axial profile anticipated above, and the reason the uniform-internal-field ellipsoid of Eq.~(\ref{e-depol}) must be read as a volume-integrated result rather than combined with a local one. No mesh-convergence study, port normalisation, or comparison of simulated against measured $f$ or $Q$ was performed, and the graphite conductivity was assumed rather than measured, so the calculation is used for the field and loss topology only.

The eigenfrequency formulation is the route by which such a calculation would eventually be tied to the measured resonance of Sec.~\ref{s-freq}. With a time dependence $e^{i\omega t}$, a complex eigenvalue $f = f' + if''$ describes a field decaying as $e^{-2\pi f''t}$, so that $Q = f'/(2f'')$ and $\Delta f = f'/Q$. Because the boundaries are open and the media absorptive the problem is non-Hermitian, and its solutions are quasinormal modes with complex frequencies~\cite{jackson1999,pozar2012,lalanne2018,sauvan2013} in which $f''$ combines radiation leakage through the perfectly matched layers~\cite{berenger1994} with absorption in the media. Since $f''$ converges considerably more slowly than $f'$ under mesh refinement, a simulated $Q$ carries larger numerical uncertainty than a simulated resonance frequency, which is why the eigenvalues ofFig.~\ref{f-simu} are quoted but not compared against a measured linewidth.

\begin{figure*}[t]
\centering\includegraphics[width=0.90\textwidth]{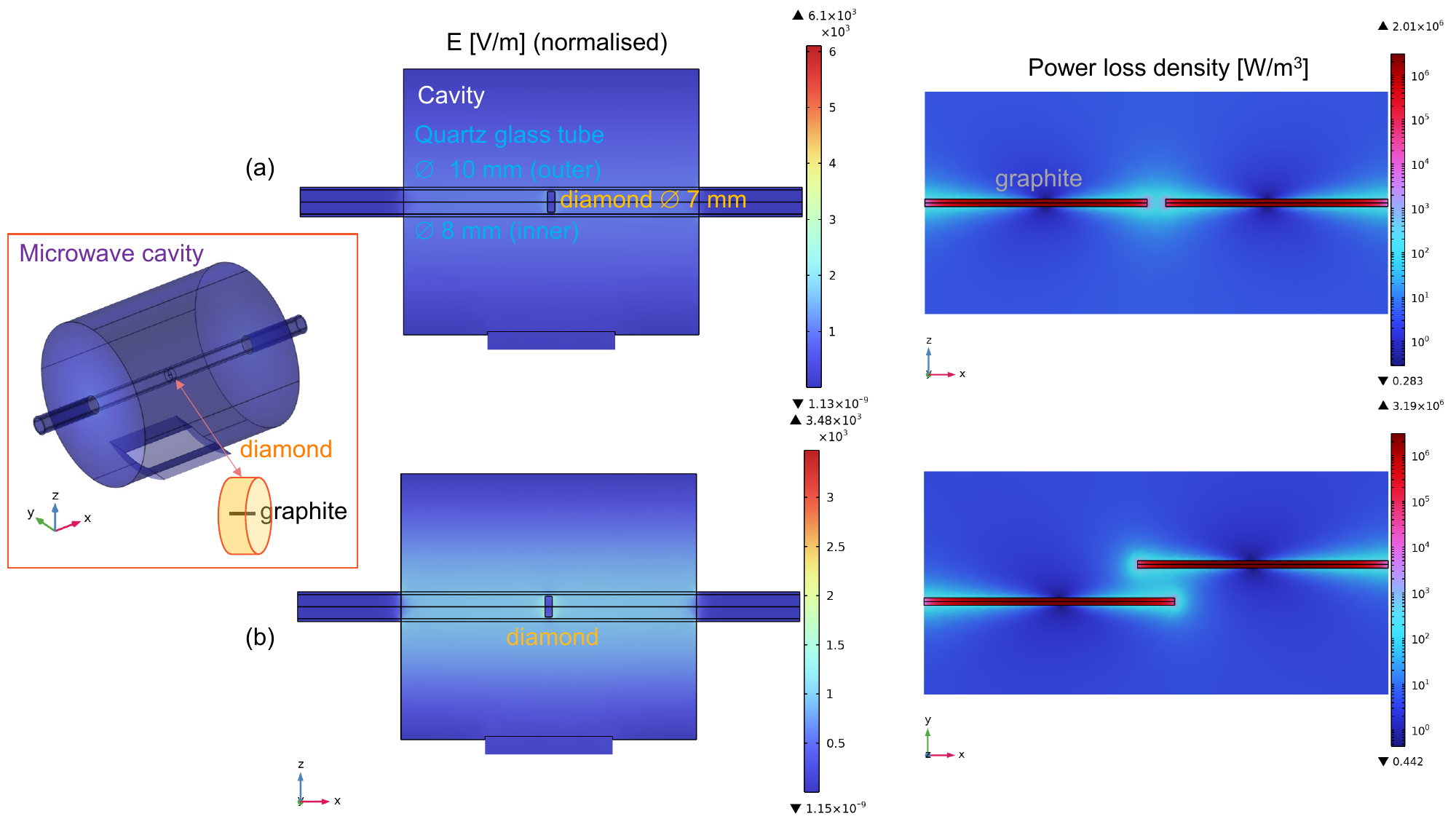}
\caption{\label{f-depo} %Electric field and volumetric power loss density $p$ (in \si{W/m^3}) for the same model as Fig.~\ref{f-simu}: graphitic elements of \SI{20}{\micro\meter} diameter inside diamond, with relative permittivity $5.5-0.001i$ and conductivity \SI{1e-12}{S/m} (diamond), and relative permittivity 10 and conductivity \SI{1000}{S/m} (graphite). The field is largest at the element terminations and across the gap, whereas the loss density peaks toward the element centres.
Electric field and volumetric power loss density for the same model and material properties as Fig. 6. Inset: 3D view of the microwave cavity, with the diamond disk (Ø 7 mm) held on the cavity axis inside a quartz glass tube (outer/inner Ø 10/8 mm). (a) Collinear graphite wires with a 0.1 mm axial gap; (b) wires staggered by a 0.1 mm lateral offset. Left: |E| in the xz-plane through the cavity axis; Right: power loss density in W/m³ around the graphite wires, in the xz/xy-plane under the same input power. While the field is highest at the wire terminations and across the gap (Fig. 6), dissipation is concentrated in the graphite and peaks toward the center of each wire.}
\end{figure*}

Figure~\ref{f-depo} resolves the same solution into the applied field and the volumetric loss density $p$ of Eq.~(\ref{e-power}), in \si{W/m^3}, and three of its features carry directly into the analysis that follows. First, the dissipation is confined to the graphitic volume: with the surrounding diamond at $\sigma = \SI{1e-12}{S/m}$ and $\varepsilon'' = 0.001$ its loss density is orders of magnitude below the graphitic value, so the crystal is heated almost entirely by conduction outward from a source occupying a $\sim\!10^{-4}$ fraction of its volume --- the localised-source geometry that Sec.~\ref{s-thermal} then evaluates thermally. Second, the two maps are complementary rather than redundant: $|E|$ is largest at the element terminations and across the gap, whereas $p$ peaks toward the element centres, following the induced current density, so the field maximum is precisely \emph{not} where the heating occurs. Third, $p = \sigma|E_{\mathrm{in}}|^{2}$ depends on how far the internal field has been screened, which is the quantity $S$ measures; the map is therefore the spatially resolved counterpart of the single number $p_{\mathrm{rel}} = 1/(1+S^{2})$ tabulated in Table~\ref{t-elements}, and the reason a track sitting near $S \approx 1$ dissipates so much more than one driven transversely. Because the modelled conductivity and geometry are simplified, the values on the colour scale should be read as indicative of the loss topology rather than as a prediction of absorbed power.

%____________________________fig 6
%____________________________fig 7

\subsection{Why depth and not diameter: the central line carries the effect}
\label{s-geometry}

The structure comprises a central axial line and a set of concentric rings, both continuous, spanning
$h = \SI{1.8}{mm}$ in depth with observed track diameter $a \approx \SI{7.5}{\micro\meter}$. The two components present very different shapes to the field and must be treated separately.

The central line is a slender rod of aspect ratio $m = h/a \approx 240$, for which Eq.~(\ref{e-L})
gives $L_\parallel \approx 9.0\times10^{-5}$ and $L_\perp \approx 0.5$. Taking
$\sigma = \SI{5000}{S/m}$ ($\rho = 0.02~\Omega\cdot$cm, the value reported for picosecond-written
wires~\cite{jedrkiewicz2017,kuriakose2024}) gives $S_\parallel \approx 0.58$ and
$S_\perp \approx 3.2\times10^{3}$. The axial value sits close to the crossover, so the field
penetrates to $|E_{\mathrm{in}}| \approx 0.87\,E_0$, while transversely it is admitted only to
$\sim\!3\times10^{-4}E_0$; the loss-density ratio for this element alone is
$(1+S_\perp^2)/(1+S_\parallel^2) \approx 8\times10^{6}$.

The rings behave quite differently. Because successive passes merge, each ring is not a slender wire but a thin-walled cylindrical shell of height $h = \SI{1.8}{mm}$ and diameter $D$ from
$\sim$\SI{0.2}{mm} to $\sim$\SI{2}{mm}, so its aspect ratio along the depth axis is
$h/D \approx 9$ down to $0.9$ --- of order unity rather than of order $10^{2}$. Table~\ref{t-elements} gives the resulting factors. For the rings $L_\parallel$ and $L_\perp$ are comparable, both of order $0.1$--$0.5$, so the rings provide almost no orientation contrast; for the outermost, squattest shell the ratio even falls slightly below unity. Both values moreover place the rings far out on the over-screened branch, $S \sim 10^{2}$--$10^{3}$, where $p \propto 1/S^{2}$ and the loss density is four to six orders of magnitude below its optimum. The rings are weak absorbers in \emph{both} orientations.

\begin{table}[t]
\caption{\label{t-elements} Depolarisation factors and screening parameters for the two components of the inscribed structure, for $h = \SI{1.8}{mm}$, $a = \SI{7.5}{\micro\meter}$ and
$\sigma = \SI{5000}{S/m}$. $p_{\mathrm{rel}} = 1/(1+S^2)$ is the loss density relative to its value at the crossover.}
\begin{ruledtabular}
\begin{tabular}{lcccccc}
Element & $h/D$ & $L_\parallel$ & $S_\parallel$ & $p_{\mathrm{rel}}^{\parallel}$ & $S_\perp$ & ratio \\
\colrule
Central line & 240 & $9.0\times10^{-5}$ & 0.58 & $7.5\times10^{-1}$ & 3218 & $7.8\times10^{6}$ \\
Ring, $D = \SI{0.2}{mm}$ & 9.0 & 0.024 & 153 & $4.3\times10^{-5}$ & 3141 & $4.2\times10^{2}$ \\
Ring, $D = \SI{0.6}{mm}$ & 3.0 & 0.109 & 700 & $2.0\times10^{-6}$ & 2868 & $1.7\times10^{1}$ \\
Ring, $D = \SI{1.0}{mm}$ & 1.8 & 0.194 & 1249 & $6.4\times10^{-7}$ & 2593 & 4.3 \\
Ring, $D = \SI{2.0}{mm}$ & 0.9 & 0.362 & 2329 & $1.8\times10^{-7}$ & 2054 & 0.8 \\
\end{tabular}
\end{ruledtabular}
\end{table}

Weighting each element by its volume, the central line --- despite occupying some two orders of
magnitude less volume than the ring system --- accounts for approximately \SI{99}{\percent} of the predicted absorption in the aligned orientation, because its loss density exceeds that of the rings by four to six orders of magnitude. The rings dominate the transverse orientation, where the central line is strongly screened, and so set a weak orientation-insensitive floor. The predicted contrast
for the structure as a whole is therefore
\begin{equation}
\frac{P_\parallel}{P_\perp} \approx 1\times10^{3},
\label{e-contrast}
\end{equation}
rather than the $10^{6}$ that the central line would give in isolation. This is large enough to account for a strong response in one orientation and nothing above a $40^\circ$C floor in the other (Sec.~\ref{s-orient}), and it explains why the depth of the structure and not its diameter is the relevant length: the depth is the long axis of the one element with an aspect ratio large enough to reach the crossover.

The model therefore makes a sharp and inexpensive prediction. The orientation selectivity should be
carried almost entirely by the central axial line: a specimen written with the central line only should show essentially the full effect, and a specimen written with rings only should show a weak response with little orientation preference. That the central line is also the most strongly graphitised feature in the infrared maps (Sec.~\ref{s-spectro}) is consistent with this assignment.
The same algebra also fixes the sensitivity of the prediction to structure, which is why continuity matters so much: were the central column fragmented into \SI{50}{\micro\meter} segments, its aspect ratio would fall to $m \approx 6.7$, $L_\parallel$ would rise to $3.7\times10^{-2}$, $S_\parallel$ to $\sim$240, and the single-element contrast would collapse from $8\times10^{6}$ to $\sim$$2\times10^{2}$. %Cross-sectional verification of continuity is correspondingly the measurement that most tightens Eq.~(\ref{e-contrast}), and Sec.~\ref{s-scope} returns to it.

\subsection{A conductivity optimum, and the direction of feedback}
\label{s-optimum}

Since $S \propto \sigma$, Eq.~(\ref{e-S}) is non-monotonic in conductivity: the loss density rises
linearly while screening is incomplete, falls as $1/\sigma$ once the inclusion screens itself, and is maximised at the crossover $S = 1$,
\begin{equation}
\sigma_{\mathrm{opt}} = \frac{\omega\varepsilon_0\varepsilon_h}{L_\parallel},
\qquad
p_{\mathrm{max}} = \frac{\omega\varepsilon_0\varepsilon_h}{2L_\parallel}\,|E_0|^{2},
\qquad
S = \frac{\sigma}{\sigma_{\mathrm{opt}}}.
\label{e-sopt}
\end{equation}
This is a depolarisation-loss maximum, not impedance matching: it is a maximum of local loss density
at \emph{fixed applied field}, whereas true matching concerns the input impedance of the loaded
cavity and the net power transferred from the generator. The distinction limits what follows,
because in a resonant cavity $E_0$ is not an independent parameter --- it depends on the loaded $Q$, the coupling coefficient, the detuning and the tuner state, all of which respond to the sample conductivity --- so maximising $p(\sigma)$ at fixed $E_0$ does not by itself identify the conductivity that maximises absorbed power at fixed \emph{launched} power.

Within that reading, the numbers are striking. For the central line
($L_\parallel \approx 9.0\times10^{-5}$) the optimum lies at $\sigma_{\mathrm{opt}} \approx \SI{8.6e3}{S/m}$, i.e.\ $\rho_{\mathrm{opt}} \approx 0.012~\Omega\cdot$cm --- squarely inside the \SIrange{0.01}{0.02}{\ohm\cdot cm} range that
laser-written graphitic wires in diamond actually reach~\cite{jedrkiewicz2017,kuriakose2024,%
momgaudis2025}. A slender laser-written track in diamond is thus close to the best possible
\SI{2.45}{GHz} absorber that its own geometry allows, which is why a structure occupying a
$10^{-4}$ fraction of the crystal volume can dominate its microwave response.

The optimum also fixes the sign of any feedback during heating. From Eq.~(\ref{e-S}),
\begin{equation}
\frac{\mathrm{d}p}{\mathrm{d}\sigma} \propto \frac{1-S^{2}}{(1+S^{2})^{2}},
\label{e-dpds}
\end{equation}
so $\mathrm{d}p/\mathrm{d}\sigma > 0$ for $\sigma < \sigma_{\mathrm{opt}}$ and
$\mathrm{d}p/\mathrm{d}\sigma < 0$ above it. If annealing raises the conductivity of the written
material, as ordering of the sp$^2$ network would, then a track starting \emph{below} the optimum
experiences positive feedback until the optimum is crossed, while one starting \emph{above} it
experiences immediate negative feedback. The stable, self-limiting branch is therefore the
over-conductive one, $\sigma > \sigma_{\mathrm{opt}}$ or equivalently $\rho < \rho_{\mathrm{opt}}$.
With $S_\parallel \approx 0.58$, the present specimen began on the unstable side, which is
qualitatively what the abrupt transition of Fig.~\ref{f-hta}(a) looks like: a smooth staircase while
the tracks climb toward $\sigma_{\mathrm{opt}}$, then a rapid excursion once positive feedback
outruns the power steps. The same argument bounds the excursion, since several processes expected at
high temperature move $\sigma$ back away from $\sigma_{\mathrm{opt}}$ --- oxidative consumption and
fragmentation of the tracks lower $\sigma$ and simultaneously reduce the aspect ratio, raising
$L_\parallel$ and lowering $\sigma_{\mathrm{opt}}$ itself. The reversal of $\Delta f$ at the
transition (Sec.~\ref{s-freq}) is consistent with such a change in the character of the load, though
the run was stopped by the operator rather than allowed to reach a terminal state, and the
contributions that these data do not separate are set out in Sec.~\ref{s-scope}.

\subsection{Thermal topology: local heating in a nearly isothermal crystal}
\label{s-thermal}

With $k \approx \SI{2000}{W.m^{-1}.K^{-1}}$ and volumetric heat capacity
$C_V \approx \SI{1.8e6}{J.m^{-3}.K^{-1}}$, the thermal diffusivity is
$\alpha = k/C_V \approx \SI{1.1e-3}{m^2/s}$ and the diffusion time across the sample thickness
$d = \SI{2}{mm}$ is $t_{\mathrm{eq}} \simeq d^{2}/4\alpha \approx \SI{1}{ms}$ --- five orders of
magnitude shorter than the minute-scale power steps. A short diffusion time means that a steady state
is reached quickly, not that it is uniform: under continuous localised deposition a stationary
gradient is maintained by the balance between generation and conduction. The relevant estimate is the
steady overheat of a localised source, $\Delta T \approx P/(4\pi k a_{\mathrm{s}})$ for a source of
characteristic radius $a_{\mathrm{s}}$ dissipating power $P$. Taking
$k = \SI{600}{W.m^{-1}.K^{-1}}$ as more representative of a nitrogen-doped, laser-damaged crystal near
$1000^\circ$C, and $P = \SI{1}{W}$, gives $\Delta T \approx \SI{27}{K}$ for
$a_{\mathrm{s}} = \SI{5}{\micro\meter}$, \SI{2.7}{K} at \SI{50}{\micro\meter} and \SI{0.3}{K} at
\SI{500}{\micro\meter}.

The picture is therefore of tracks running a few tens of kelvin above their surroundings inside a
crystal that is close to isothermal on millimetre scales. Both halves matter. The local overheat is
what allows a $10^{-4}$ volume fraction to drive the oxidation chemistry observed at the surfaces
(Sec.~\ref{s-morph}), while the near-uniformity on the scale of the specimen means that the
polarisation selectivity of Sec.~\ref{s-orient} selects where power is \emph{deposited} and not which region of the crystal receives a thermal \emph{dose}. Field-addressed spatially selective annealing of a millimetre-sized diamond over minute timescales is thus excluded by the host's own thermal conductivity; it would require pulses shorter than the diffusion time, membrane or nanoscale geometries, thermal barriers, or a lower-conductivity host. Thermal uniformity also carries no information about where the power goes: a body of high thermal conductivity can be nearly isothermal while all of its power is deposited in a very small fraction of its volume.

\subsection{The nitrogen donor as a second channel at high temperature}
\label{s-nitrogen}

Thermal ionisation of the deep nitrogen donor offers a candidate second absorption channel once the crystal is hot. Equilibrium statistics~\cite{sze2007,farrer1969} give
\begin{equation}
n \simeq \sqrt{\tfrac{1}{2}N_c N_s}\;\exp\!\left(-\frac{E_d}{2k_BT}\right),
\label{e-carriers}
\end{equation}
with $n$ the free-electron concentration, $N_c \sim \SI{e20}{cm^{-3}}$ the effective conduction-band density of states, $N_s = \SI{1.8e19}{cm^{-3}}$, $E_d = \SI{1.7}{eV}$, $k_B$ the Boltzmann constant and $T$ the absolute temperature. This yields $n \sim \SI{e5}{cm^{-3}}$ at \SI{300}{K} and
$n \sim \SI{e16}{cm^{-3}}$ at $800^\circ$C; with an electron mobility $\mu_e$ falling from
$\sim\!10^{3}$ to $\sim\!10^{2}$~cm$^2$V$^{-1}$s$^{-1}$, the bulk conductivity $\sigma = q n \mu_e$ would rise by roughly ten orders of magnitude across that interval. A host that is transparent to the field when cold can therefore become weakly lossy in its own right once the inscribed tracks have raised its temperature, which would make the inscription an igniter for a volumetric channel rather than the whole absorber.

We keep this as an order-of-magnitude estimate rather than a component of the argument. Equation~(\ref{e-carriers}) assumes an uncompensated single-donor system, whereas the Ni-related charge transfer noted in Sec.~\ref{s-exp} converts N$_s^0$ to N$_s^+$ by an unmeasured amount; $N_c$ and $\mu_e$ are not constrained for this defective crystal. Nothing in the present data distinguishes this channel from continued absorption in the graphitic tracks, since the near-isothermal condition of Sec.~\ref{s-thermal} carries no information on where the power is deposited. The decisive test is a series of identically written specimens spanning several nitrogen concentrations, including type-IIa material. If the transition threshold and the cavity loss track $N_s$, this channel is real; if they do not, the graphitic tracks account for the whole response.

\subsection{What the present data fix, and what they leave open}
\label{s-scope}

The orientation selectivity of Sec.~\ref{s-orient} and the depolarisation argument of Secs.~\ref{s-S}--\ref{s-geometry} are the load-bearing content of this paper, and it is worth stating compactly what they rest on. Two inputs come from outside this experiment. The conductivity of the written tracks is adopted from the picosecond-writing literature~\cite{jedrkiewicz2017,kuriakose2024}, whereas these structures were written with \SI{200}{fs} pulses, a regime associated with more disordered and less conductive carbon; laser-written carbon is moreover heterogeneous and dispersive, so a scalar conductivity may be an inadequate constitutive parameter at \SI{2.45}{GHz}. This input sets the absolute values of $S$ and hence which side of $\sigma_{\mathrm{opt}}$ the specimen begins on. The second input is importance of the axial continuity, inferred from the writing conditions and the optical micrographs. %rather than demonstrated by cross-sectional Raman mapping, FIB sectioning or an electrical measurement; Eq.~(\ref{e-contrast}) moves by five orders of magnitude if it fails. 
A resistance measurement across the \SI{1.8}{mm} depth of a witness column written under identical conditions would supply %both at once, and is 
the single most valuable measurement outstanding.

Three contributions to the \emph{magnitude} of the transition are not separated by these data. The thermometry is a single-band instrument viewing an optically evolving target, so an emissivity change and a temperature change enter it identically: the indicated excursion from 500 to $900^\circ$C is compatible with a true excursion of only $\sim$\SI{160}{K} accompanied by an emissivity rise from 0.1 to 0.8, as quantified in Appendix~\ref{a-emis}, and two-colour pyrometry would resolve it. Back-side ablation is visible in Fig.~\ref{f-setups}(a), so surface graphite may contribute alongside the buried tracks, and carbon oxidation is exothermic, so chemical heat release contributes in an unknown proportion to dielectric heating. Beyond these, only launched power is reported: the external area of the crystal is \SI{64}{mm^2} and even a blackbody at $900^\circ$C radiates about \SI{6.8}{W} above room-temperature surroundings, so the power actually absorbed is likely of order a few watts and \SI{90}{W} is a generator setting rather than a measure of coupling efficiency. % in the most of experiments, the reflected power was below 5\%. %Synchronous logging of forward and reflected power, $S_{11}$ and loaded $Q$ would close that balance and test whether the transition is partly a retuning of an actively matched cavity rather than a change in the sample.
Reflected power remained below 5\% throughout the experiment. During the final power increment, both the resonant frequency and the quality factor decreased (Fig.~\ref{f-Q}). The reduction in $Q$ indicates increased microwave losses, while the simultaneous frequency shift suggests a change in the electromagnetic loading of the cavity. Together, these observations are consistent with a temperature-induced transition in the coupling between the laser-written graphitic structure and the microwave field, marking the onset of a different cavity-loading regime rather than a response solely attributable to geometrical changes associated with oxidation.

%None of this touches the orientation contrast itself, which is a comparison between two configurations\ of the same instrument and the same crystal. And the sharpest test of the model requires no new instrumentation: writing one specimen with the central line only and one with rings only should, on the argument of Sec.~\ref{s-geometry}, reproduce and suppress the effect respectively.

\section{Conclusions and Outlook}
\label{s-conclusions}

Femtosecond-laser inscription converts a microwave-transparent diamond into a microwave absorber, and the conversion is directional. A $4\times4\times2$~mm$^3$ N-doped HPHT crystal carrying a buried graphitised concentric grating, driven at \SI{2.45}{GHz}, produced a stepwise pyrometric response through 5--\SI{10}{W} power increments to an indicated $\sim 500^\circ$C at \SI{80}{W} and then an
abrupt excursion to an indicated $\sim 900^\circ$C at \SI{90}{W}, accompanied by surface roughening and terraced oxidation; the same crystal rotated by $90^\circ$, and an unmodified crystal in either orientation, stayed below the $40^\circ$C detection floor. The effect belongs neither to diamond nor
to laser-modified diamond as such, but to the modified crystal in a particular relation to the field.

The orientation dependence follows quantitatively from the depolarisation factors of the inscribed sub-wavelength tracks, which collapse geometry, material and drive into the single screening parameter $S = L\sigma/(\omega\varepsilon_0\varepsilon_h) = \sigma/\sigma_{\mathrm{opt}}$. Treating the
continuous central line and the continuous ring shells separately gives a predicted contrast of order $10^{3}$ for the structure as a whole, carried almost entirely by the central line, whose aspect ratio of $\sim$240 places it near the crossover $S \approx 1$ while the ring shells, of aspect ratio of order unity, sit far out on the over-screened branch in both orientations. The picture is internally consistent with the infrared maps, which find the strongest sp$^2$ signature on the central line, and it yields a test that costs one writing run: centre-line-only structures should reproduce the effect, rings-only structures should largely suppress it.

Three consequences follow for the design of laser-written microwave absorbers. First, every term in $S$ is a writing parameter or an instrument setting, element length and diameter set $L$, the writing regime sets $\sigma$, and the source sets $\omega$, so the model is a design rule and not only an explanation. Second, the optimum $\rho_{\mathrm{opt}} \approx 0.012~\Omega\cdot$cm for a \SI{1.8}{mm} slender track falls inside the range that laser writing already achieves, and the
over-conductive side of the optimum is the stable one, which makes writing more conductive than $\rho_{\mathrm{opt}}$ the robust design choice. Third, since nothing in
Eqs.~(\ref{e-depol})-(\ref{e-sopt}) is specific to carbon or to diamond, the same algebra transfers
to any low-loss host in which conductive tracks can be written: SiC, sapphire, silica.  The scaling $S \propto 1/\omega$ makes the approach less favourable at industrial RF frequencies and more favourable toward millimetre-wave sources, which can be used to create temperature gradients in substrates during film growth.

Two boundaries should be stated as plainly as the result. The host equilibrates thermally in $\sim$\SI{1}{ms}, so polarisation selects where power is deposited but not which region receives a thermal dose: spatially selective annealing of a millimetre-sized diamond is not achievable this way, and no claim is made about NV-centre processing, where the obstacle here is uncontrolled surface oxidation rather than survival of the centres. And the quantitative side of the model rests on a borrowed conductivity and an inferred continuity, with a single-band pyrometer as the only thermometer; the measurements that would remove those dependencies are specific, inexpensive and set out in Sec.~\ref{s-scope}. What stands without them is an orientation-selective microwave absorber whose active element can be placed anywhere inside a transparent crystal by a laser, together with a closed-form account of why its orientation matters and how strong it can be made.

\small\begin{acknowledgments}
M. Z. acknowledges support via MEXT Next-Generation Computational Science Grand Reach Program (Grant No.JPMXP1920260105). S. J. acknowledges support via the Australian Research Council DP240103231 grant.
%J.M. acknowledges support via JST CREST (Grant No. JPMJCR19I3) and KAKENHI (No.22H02137). 
J. M. acknowledges support via JST CREST (Grant No. JPMJCR19I3) and KAKENHI (No.26K01228). M. R. acknowledges support via JST-PRESTO (JPMJPR250E).

\end{acknowledgments}

\bibliography{aipsamp%,v2extra,v4extra
}

@PREAMBLE{
 "\providecommand{\noopsort}[1]{}" 
 # "\providecommand{\singleletter}[1]{#1}%" 
}

@article{Massi,
title = {Microwave-assisted dehydration of calcium hydroxide for thermochemical energy storage},
journal = {Journal of Energy Storage},
volume = {108},
pages = {115195},
year = {2025},
issn = {2352-152X},
doi = {https://doi.org/10.1016/j.est.2024.115195},
url = {https://www.sciencedirect.com/science/article/pii/S2352152X24047819},
author = {Massimiliano Zamengo and Hisahiro Einaga and Yuji Wada and Junko Morikawa}
}

@article{16semsc221,
title = {Light-induced reflectivity transients in black-Si nanoneedles},
journal = {Solar Energy Materials and Solar Cells},
volume = {144},
pages = {221-227},
year = {2016},
author = {P. Ščajev and T. Malinauskas and G. Seniutinas and M.D. Arnold and A. Gentle and I. Aharonovich and G. Gervinskas and P. Michaux and J.S. Hartley and E.L.H. Mayes and P.R. Stoddart and S. Juodkazis}
}

@Inbook{Horikoshi2024,
author="Horikoshi, Satoshi
and Catal{\'a}-Civera, Jos{\'e} M.
and Schiffmann, Robert F.
and Fukushima, Jun
and Mitani, Tomohiko
and Serpone, Nick",
title="Microwave Heating",
bookTitle="Microwave Chemical and Materials Processing: A Tutorial",
year="2024",
publisher="Springer Nature Singapore",
address="Singapore",
pages="53--93",
isbn="978-981-97-5795-4",
doi="10.1007/978-981-97-5795-4_4",
url="https://doi.org/10.1007/978-981-97-5795-4_4"
}

@ARTICLE{Osep,
  author={Osepchuk, J.M.},
  journal={IEEE Transactions on Microwave Theory and Techniques}, 
  title={A History of Microwave Heating Applications}, 
  year={1984},
  volume={32},
  number={9},
  pages={1200-1224},
  doi={10.1109/TMTT.1984.1132831}}

@article{Truong2025,
  author  = {T Truong and W Liang and R Basnet and W Nemeth and P Stradins and DL Young and D Macdonald and KC Fong},
  title   = {Microwave annealing for fast and effective hydrogen activation in polycrystalline silicon passivating contacts},
  journal = {Advanced Energy and Sustainability Research},
  year    = {2025},
  volume = {6},
    number = {10},
    pages = {2500004},
  doi     = {10.1002/aesr.202500004}
}

@article{sutton1989,
  author  = {Sutton, Willard H.},
  title   = {Microwave processing of ceramic materials},
  journal = {American Ceramic Society Bulletin},
  volume  = {68},
  pages   = {376--386},
  year    = {1989}
}

@incollection{janney1988,
  author    = {Janney, Mark A. and Kimrey, Hal D.},
  title     = {Microwave sintering of alumina at 28 {GHz}},
  booktitle = {Ceramic Transactions, Vol.~1: Ceramic Powder Science IIB},
  publisher = {American Ceramic Society},
  pages     = {919--924},
  year      = {1988}
}

@inproceedings{janney1990,
  author    = {Janney, Mark A. and Kimrey, Hal D.},
  title     = {Diffusion-controlled processes in microwave-fired oxide ceramics},
  booktitle = {Materials Research Society Symposium Proceedings},
  volume    = {189},
  pages     = {215--227},
  year      = {1990}
}

@article{janney1991,
  author  = {Janney, Mark A. and Kimrey, Hal D. and Allen, W. R. and Kiggans, James O.},
  title   = {Grain growth in microwave-annealed alumina},
  journal = {Journal of the American Ceramic Society},
  volume  = {74},
  year    = {1991},
  doi     = {10.1111/j.1151-2916.1991.tb07159.x}
}

@article{brosnan2003,
  author  = {Brosnan, Kristen H. and Messing, Gary L. and Agrawal, Dinesh K.},
  title   = {Microwave sintering of alumina at 2.45 {GHz}},
  journal = {Journal of the American Ceramic Society},
  volume  = {86},
  year    = {2003},
  doi     = {10.1111/j.1151-2916.2003.tb03467.x}
}

@article{rybakov1994,
  author  = {Rybakov, Kirill I. and Semenov, Vladimir E.},
  title   = {Possibility of plastic deformation of an ionic crystal due to the nonthermal influence of a high-frequency electric field},
  journal = {Physical Review B},
  volume  = {49},
  pages   = {64--68},
  year    = {1994}
}

@article{rybakov1995,
  author  = {Rybakov, Kirill I. and Semenov, Vladimir E.},
  title   = {Mass transport in ionic crystals induced by the ponderomotive action of a high-frequency electric field},
  journal = {Physical Review B},
  volume  = {52},
  pages   = {3030--3033},
  year    = {1995}
}

@article{rybakov1997,
  author  = {Rybakov, Kirill I. and Semenov, Vladimir E. and Freeman, S. A. and Booske, John H. and Cooper, Reid F.},
  title   = {Dynamics of microwave-induced currents in ionic crystals},
  journal = {Physical Review B},
  volume  = {55},
  pages   = {3559--3567},
  year    = {1997}
}

@article{booske1997,
  author  = {Booske, John H. and Cooper, Reid F. and Freeman, S. A.},
  title   = {Mechanisms for nonthermal effects on ionic mobility during microwave processing of crystalline solids},
  journal = {Journal of Materials Research},
  year    = {1997}
}

@article{bykov2001jpd,
  author  = {Bykov, Yu.~V. and Rybakov, Kirill I. and Semenov, Vladimir E.},
  title   = {High-temperature microwave processing of materials},
  journal = {Journal of Physics D: Applied Physics},
  volume  = {34},
  pages   = {R55--R75},
  year    = {2001}
}

@article{srep2016,
  title   = {The effect of electric field intensification at interparticle contacts in microwave sintering},
  journal = {Scientific Reports},
  volume  = {6},
  pages   = {32163},
  year    = {2016}
}

@article{roy1999,
  author  = {Roy, Rustum and Agrawal, Dinesh and Cheng, Jiping and Gedevanishvili, Shalva},
  title   = {Full sintering of powdered-metal bodies in a microwave field},
  journal = {Nature},
  volume  = {399},
  number  = {6737},
  pages   = {668--670},
  year    = {1999},
  doi     = {10.1038/21390}
}

@article{bhattacharya2016,
  author  = {Bhattacharya, M. and Basak, T.},
  title   = {A review on the susceptor assisted microwave processing of materials},
  journal = {Energy},
  volume  = {97},
  pages   = {306--338},
  year    = {2016}
}

@article{sic2007,
  title   = {Ultrahigh-temperature microwave annealing of {Al$^+$}- and {P$^+$}-implanted {4H-SiC}},
  journal = {Journal of Applied Physics},
  year    = {2007}
}

@article{lowtemp2013,
  title   = {Effective dopant activation via low temperature microwave annealing of ion implanted silicon},
  journal = {Applied Physics Letters},
  year    = {2013}
}

@article{susceptor2012,
  title   = {Susceptor coupling for the uniformity and dopant activation efficiency in implanted {Si} under fixed-frequency microwave anneal},
  journal = {IEEE Electron Device Letters},
  year    = {2012}
}

@article{carbon2014,
  title   = {Activation of silicon implanted with phosphorus and boron atoms by microwave annealing with carbon powder as a heat source},
  journal = {Japanese Journal of Applied Physics},
  year    = {2014}
}

@article{ge_apl2016,
  title   = {High dopant activation of phosphorus in {Ge} crystal with high-temperature implantation and two-step microwave annealing},
  journal = {Applied Physics Letters},
  year    = {2016}
}

@article{geonsi2016,
  title   = {Increased photoluminescence from single-mode microwave annealing of n-type {Ge-on-Si}},
  journal = {Applied Physics Letters},
  year    = {2016}
}

@article{nanosheet2022,
  title   = {Efficient and stable activation by microwave annealing of nanosheet silicon doped with phosphorus above its solubility limit},
  journal = {Applied Physics Letters},
  year    = {2022},
  doi     = {10.1063/5.0099083}
}

@article{mechanistic2025,
  title   = {Mechanistic insights into microwave annealing for lattice defect recovery},
  journal = {IEEE Journal of the Electron Devices Society},
  year    = {2025}
}

@article{nonthermal_review2025,
  title   = {On the non-thermal mechanisms in microwave sintering of materials},
  journal = {Materials},
  volume  = {18},
  pages   = {668},
  year    = {2025},
  doi     = {10.3390/ma18030668}
}

@article{jedrkiewicz2017,
  author  = {Kumar, S. and Sotillo, B. and Chiappini, A. and Ramponi, R. and
             Di Trapani, P. and Eaton, S. M. and Jedrkiewicz, O.},
  title   = {Study of graphitic microstructure formation in diamond bulk by
             pulsed {Bessel} beam laser writing},
  journal = {Applied Physics A},
  volume  = {123},
  pages   = {698},
  year    = {2017},
  doi     = {10.1007/s00339-017-1303-9}
}

@article{kuriakose2023,
  author  = {Kuriakose, A. and Mezzapesa, F. P. and Gaudiuso, C. and Chiappini, A. and
             Ancona, A. and Jedrkiewicz, O.},
  title   = {Fabrication of conductive micro electrodes in diamond bulk using
             pulsed {Bessel} beams},
  journal = {Diamond and Related Materials},
  year    = {2023},
  doi     = {10.1016/j.diamond.2023.110139}
}

@article{kuriakose2024,
  author  = {Kuriakose, A. and Mezzapesa, F. P. and Gaudiuso, C. and Chiappini, A. and
             Picollo, F. and Ancona, A. and Jedrkiewicz, O.},
  title   = {{Bessel} beam fabrication of graphitic micro electrodes in diamond
             using laser bursts},
  journal = {Diamond and Related Materials},
  volume  = {147},
  pages   = {111316},
  year    = {2024},
  doi     = {10.1016/j.diamond.2024.111316}
}

@article{momgaudis2025,
  author  = {Momgaudis, B. and Clerici, M. and Jukna, V. and Jedrkiewicz, O. and others},
  title   = {Dynamics of {Bessel} beam induced graphitization of diamond:
             exploring the role of in-bulk plasma},
  journal = {Optical Materials Express},
  year    = {2025}
}

@article{sotillo2016,
  author  = {Sotillo, B. and Bharadwaj, V. and Hadden, J. P. and Sakakura, M. and
             Chiappini, A. and Fernandez, T. T. and Longhi, S. and Jedrkiewicz, O. and
             Shimotsuma, Y. and Criante, L. and Osellame, R. and Galzerano, G. and
             Ferrari, M. and Miura, K. and Ramponi, R. and Barclay, P. E. and Eaton, S. M.},
  title   = {Diamond photonics platform enabled by femtosecond laser writing},
  journal = {Scientific Reports},
  volume  = {6},
  pages   = {35566},
  year    = {2016},
  doi     = {10.1038/srep35566}
}

@article{kononenko2017,
  author  = {Kononenko, V. V. and Khomich, A. A. and Konov, V. I.},
  title   = {Peculiarities of laser-induced material transformation inside diamond bulk},
  journal = {Diamond and Related Materials},
  year    = {2017}
}

@article{ferrari2000,
  author  = {Ferrari, A. C. and Robertson, J.},
  title   = {Interpretation of {Raman} spectra of disordered and amorphous carbon},
  journal = {Physical Review B},
  volume  = {61},
  pages   = {14095--14107},
  year    = {2000},
  doi     = {10.1103/PhysRevB.61.14095}
}

@article{blackdiamond2019,
  title   = {Electrical conductivity of double textured black diamond films from
             RT to 800 K},
  journal = {Diamond and Related Materials},
  year    = {2019},
  doi     = {10.1016/j.diamond.2018.12.005}
}

@article{carbonado1998,
  author  = {Heaney, P. J. and Vicenzi, E. P. and De, S.},
  title   = {Microstructural observations of polycrystalline diamond: a contribution
             to the carbonado conundrum},
  journal = {Earth and Planetary Science Letters},
  year    = {1998},
  doi     = {10.1016/S0012-821X(98)00229-5}
}

@article{nvimplant2017,
  author  = {Prawer, S. and others},
  title   = {An overview on the formation and processing of nitrogen-vacancy photonic
             centers in diamond by ion implantation},
  journal = {Journal of Manufacturing and Materials Processing},
  volume  = {1},
  pages   = {6},
  year    = {2017},
  doi     = {10.3390/jmmp1010006}
}

@article{green2022rimg,
  author  = {Green, Benjamin L. and Collins, Alan T. and Breeding, Christopher M.},
  title   = {Diamond spectroscopy, defect centers, color, and treatments},
  journal = {Reviews in Mineralogy and Geochemistry},
  volume  = {88},
  number  = {1},
  pages   = {637--688},
  year    = {2022},
  doi     = {10.2138/rmg.2022.88.12}
}

@book{zaitsev2001,
  author    = {Zaitsev, Alexander M.},
  title     = {Optical Properties of Diamond: A Data Handbook},
  publisher = {Springer},
  address   = {Berlin, Heidelberg},
  year      = {2001},
  doi       = {10.1007/978-3-662-04548-0}
}

@article{yelisseyev2007,
  author  = {Yelisseyev, Alexander and Kanda, Hisao},
  title   = {Optical centers related to {3d} transition metals in diamond},
  journal = {New Diamond and Frontier Carbon Technology},
  volume  = {17},
  number  = {3},
  pages   = {127--178},
  year    = {2007}
}

@article{jones2009,
  author  = {Jones, Robert and Goss, Jonathan P. and Briddon, Patrick R.},
  title   = {Acceptor level of nitrogen in diamond and the 270-nm absorption band},
  journal = {Physical Review B},
  volume  = {80},
  number  = {3},
  pages   = {033205},
  year    = {2009},
  doi     = {10.1103/PhysRevB.80.033205}
}

@article{huang2022ns,
  author  = {Huang, Mengjiao and others},
  title   = {Rapid determination of single substitutional nitrogen {N$_s$}
             concentration in diamond from {UV--Vis} spectroscopy},
  journal = {Applied Physics Letters},
  volume  = {121},
  number  = {6},
  pages   = {064002},
  year    = {2022},
  doi     = {10.1063/5.0102757}
}

@article{davies1976nv,
  author  = {Davies, Gordon and Hamer, M. F.},
  title   = {Optical studies of the 1.945 eV vibronic band in diamond},
  journal = {Proceedings of the Royal Society of London A},
  volume  = {348},
  number  = {1653},
  pages   = {285--298},
  year    = {1976},
  doi     = {10.1098/rspa.1976.0039}
}

@article{goss1996nv,
  author  = {Goss, Jonathan P. and Jones, Robert and Breuer, S. J. and
             Briddon, Patrick R. and {\"O}berg, Sven},
  title   = {The twelve-line 1.682 eV luminescence center in diamond and the
             vacancy-silicon complex},
  journal = {Physical Review Letters},
  volume  = {77},
  number  = {14},
  pages   = {3041--3044},
  year    = {1996},
  doi     = {10.1103/PhysRevLett.77.3041}
}

@article{davies2013nv0,
  author  = {Davies, Gordon and others},
  title   = {Assignment of the {NV$^0$} 575 nm zero-phonon line in diamond to a
             {$^2$E}--{$^2$A$_2$} transition},
  journal = {arXiv preprint arXiv:1301.3542},
  year    = {2013}
}

@article{nazare1991,
  author  = {Nazar{\'e}, Maria Helena and Neves, A. J. and Davies, Gordon},
  title   = {Optical studies of the 1.40-eV {Ni} center in diamond},
  journal = {Physical Review B},
  volume  = {43},
  number  = {17},
  pages   = {14196--14205},
  year    = {1991},
  doi     = {10.1103/PhysRevB.43.14196}
}

@article{farrer1969,
  author  = {Farrer, R. G.},
  title   = {On the substitutional nitrogen donor in diamond},
  journal = {Solid State Communications},
  volume  = {7},
  number  = {9},
  pages   = {685--688},
  year    = {1969},
  doi     = {10.1016/0038-1098(69)90593-6}
}

@article{cui1998,
  author  = {Cui, J. B. and Ristein, J. and Ley, L.},
  title   = {Electron affinity of the bare and hydrogen covered single crystal
             diamond (111) surface},
  journal = {Physical Review Letters},
  volume  = {81},
  number  = {2},
  pages   = {429--432},
  year    = {1998},
  doi     = {10.1103/PhysRevLett.81.429}
}

@article{riedel2004,
  author  = {Riedel, M. and Ristein, J. and Ley, L.},
  title   = {The impact of ozone on the surface conductivity of single crystal
             diamond},
  journal = {Diamond and Related Materials},
  volume  = {13},
  number  = {4--8},
  pages   = {746--750},
  year    = {2004},
  doi     = {10.1016/j.diamond.2003.11.020}
}

@article{rutter1998,
  author  = {Rutter, M. J. and Robertson, J.},
  title   = {Ab initio calculation of electron affinities of diamond surfaces},
  journal = {Physical Review B},
  volume  = {57},
  number  = {15},
  pages   = {9241--9245},
  year    = {1998},
  doi     = {10.1103/PhysRevB.57.9241}
}

@article{ristein2000,
  author  = {Ristein, J.},
  title   = {Electronic properties of diamond surfaces --- blessing or curse for
             devices?},
  journal = {Diamond and Related Materials},
  volume  = {9},
  number  = {3--6},
  pages   = {1129--1137},
  year    = {2000},
  doi     = {10.1016/S0925-9635(99)00316-7}
}

@article{osborn1945,
  author  = {Osborn, J. A.},
  title   = {Demagnetizing factors of the general ellipsoid},
  journal = {Physical Review},
  volume  = {67},
  number  = {11--12},
  pages   = {351--357},
  year    = {1945},
  doi     = {10.1103/PhysRev.67.351}
}

@article{stoner1945,
  author  = {Stoner, E. C.},
  title   = {The demagnetizing factors for ellipsoids},
  journal = {The London, Edinburgh, and Dublin Philosophical Magazine and Journal of Science},
  volume  = {36},
  number  = {263},
  pages   = {803--821},
  year    = {1945},
  doi     = {10.1080/14786644508521510}
}

@book{bohren1983,
  author    = {Bohren, Craig F. and Huffman, Donald R.},
  title     = {Absorption and Scattering of Light by Small Particles},
  publisher = {Wiley},
  address   = {New York},
  year      = {1983},
  note      = {Ch.~5: absorption by small ellipsoids in the electrostatic approximation},
  doi       = {10.1002/9783527618156}
}

@book{sihvola1999,
  author    = {Sihvola, Ari},
  title     = {Electromagnetic Mixing Formulas and Applications},
  series    = {IEE Electromagnetic Waves Series},
  volume    = {47},
  publisher = {Institution of Electrical Engineers},
  address   = {London},
  year      = {1999},
  isbn      = {978-0-85296-772-0}
}

@book{landau1984ecm,
  author    = {Landau, L. D. and Lifshitz, E. M. and Pitaevskii, L. P.},
  title     = {Electrodynamics of Continuous Media},
  edition   = {2},
  series    = {Course of Theoretical Physics},
  volume    = {8},
  publisher = {Pergamon Press},
  address   = {Oxford},
  year      = {1984}
}

@book{sze2007,
  author    = {Sze, S. M. and Ng, Kwok K.},
  title     = {Physics of Semiconductor Devices},
  edition   = {3},
  publisher = {Wiley},
  address   = {Hoboken, NJ},
  year      = {2007},
  doi       = {10.1002/0470068329}
}

@book{jackson1999,
  author    = {Jackson, John David},
  title     = {Classical Electrodynamics},
  edition   = {3},
  publisher = {John Wiley \& Sons},
  address   = {New York},
  year      = {1999}
}

@book{pozar2012,
  author    = {Pozar, David M.},
  title     = {Microwave Engineering},
  edition   = {4},
  publisher = {John Wiley \& Sons},
  address   = {Hoboken, NJ},
  year      = {2012}
}

@article{berenger1994,
  author  = {B{\'e}renger, Jean-Pierre},
  title   = {A Perfectly Matched Layer for the Absorption of Electromagnetic Waves},
  journal = {Journal of Computational Physics},
  volume  = {114},
  number  = {2},
  pages   = {185--200},
  year    = {1994},
  doi     = {10.1006/jcph.1994.1159}
}

@article{lalanne2018,
  author  = {Lalanne, Philippe and Yan, Wei and Vynck, Kevin and Sauvan, Christophe
             and Hugonin, Jean-Paul},
  title   = {Light Interaction with Photonic and Plasmonic Resonances},
  journal = {Laser \& Photonics Reviews},
  volume  = {12},
  number  = {5},
  pages   = {1700113},
  year    = {2018},
  doi     = {10.1002/lpor.201700113}
}

@article{sauvan2013,
  author  = {Sauvan, Christophe and Hugonin, Jean-Paul and Maksymov, Ivan S.
             and Lalanne, Philippe},
  title   = {Theory of the Spontaneous Optical Emission of Nanosize Photonic
             and Plasmon Resonators},
  journal = {Physical Review Letters},
  volume  = {110},
  number  = {23},
  pages   = {237401},
  year    = {2013},
  doi     = {10.1103/PhysRevLett.110.237401}
}

@book{cbar,
  author    = {Chapman WAJ},
  title     = {Workshop technology, Part 1,},
  edition   = {5},
  publisher = {Elsevier},
  address   = {Burlington, MA},
  year      = {1972}
}

@article{13oe6901,
author = {Albertas \v{Z}ukauskas and Mangirdas Malinauskas and Ar\={u}nas Kadys and Gediminas Gervinskas and Gediminas Seniutinas and Sasikaran Kandasamy and Saulius Juodkazis},
journal = {Opt. Express},
number = {6},
pages = {6901--6909},
publisher = {Optica Publishing Group},
title = {Black silicon: substrate for laser 3D micro/nano-polymerization},
volume = {21},
month = {Mar},
year = {2013},
url = {https://opg.optica.org/oe/abstract.cfm?URI=oe-21-6-6901},
doi = {10.1364/OE.21.006901},
}

@article{RYU2020120337,
title = {Quadrupole modelling of dual lock-in method for the simultaneous measurements of thermal diffusivity and thermal effusivity},
journal = {International Journal of Heat and Mass Transfer},
volume = {162},
pages = {120337},
year = {2020},
issn = {0017-9310},
doi = {https://doi.org/10.1016/j.ijheatmasstransfer.2020.120337},
url = {https://www.sciencedirect.com/science/article/pii/S0017931020332737},
author = {Meguya Ryu and Jean-Christophe Batsale and Junko Morikawa}
}

@article{Wang2026,
  author  = {Wang, Zhuo  },
  title   = {3D lithography of diamond thermal emitters for microscale emissivity control},
  journal = {Nature Communications},
  volume  = {17},
  number  = {1},
  pages   = {8715},
  year    = {2026},
  doi     = {10.1038/s41467-026-75017-5}
}

\appendix
\setcounter{figure}{0}\setcounter{equation}{0}
\setcounter{section}{0}\setcounter{table}{0}
\renewcommand{\thefigure}{A\arabic{figure}}
\renewcommand{\thetable}{A\arabic{table}}
\renewcommand{\theequation}{A\arabic{equation}}
\renewcommand{\thesection}{A\arabic{section}}

\section{Laser writing parameters}
\label{a-writing}

Table~\ref{t-writing} collects the inscription parameters referred to in Sec.~\ref{s-exp} and
Fig.~\ref{f-setups}(a).

\begin{table}[h!]
\caption{\label{t-writing} Laser writing parameters.}
\begin{ruledtabular}
\begin{tabular}{lll}
Parameter & Symbol & Value \\
\colrule
Wavelength & $\lambda_L$ & \SI{1030}{nm} \\
Pulse duration & $t_p$ & \SI{200}{fs} \\
Repetition rate & $f_{\mathrm{rep}}$ & \SI{602.7}{kHz} \\
Numerical aperture & $\mathrm{NA}$ & 0.7 \\
Average power on sample & -- & \SI{742.9}{mW} \\
Pulse energy & $E_p$ & \SI{1.233}{\micro\joule} \\
Scan speed & $v$ & \SI{1}{mm/s} \\
Axial step between passes & $\Delta z$ & \SI{50}{\micro\meter} \\
Number of passes & -- & 40 (last 7 at reduced power) \\
Pulses per spot, central line & -- & 2000 per pass \\
Structure depth & $h$ & $\sim$\SI{1.8}{mm} \\
Track diameter (observed) & $a$ & \SIrange{5}{10}{\micro\meter} \\
\colrule
\multicolumn{3}{l}{\emph{Nominal diffraction-limited values, upper bounds only}}\\
Focal spot diameter & $2r$ & \SI{1.8}{\micro\meter} \\
Fluence & $F_p$ & \SI{48.5}{J/cm^2} \\
Irradiance & $I_p$ & \SI{242.2}{TW/cm^2} \\
Depth of focus & $2z_R$ & \SI{4.9}{\micro\meter} \\
\end{tabular}
\end{ruledtabular}
\end{table}

\section{The buried graphite--diamond boundary}
\label{a-junction}

\begin{figure*}[h!]
\centering\includegraphics[width=0.89\textwidth]{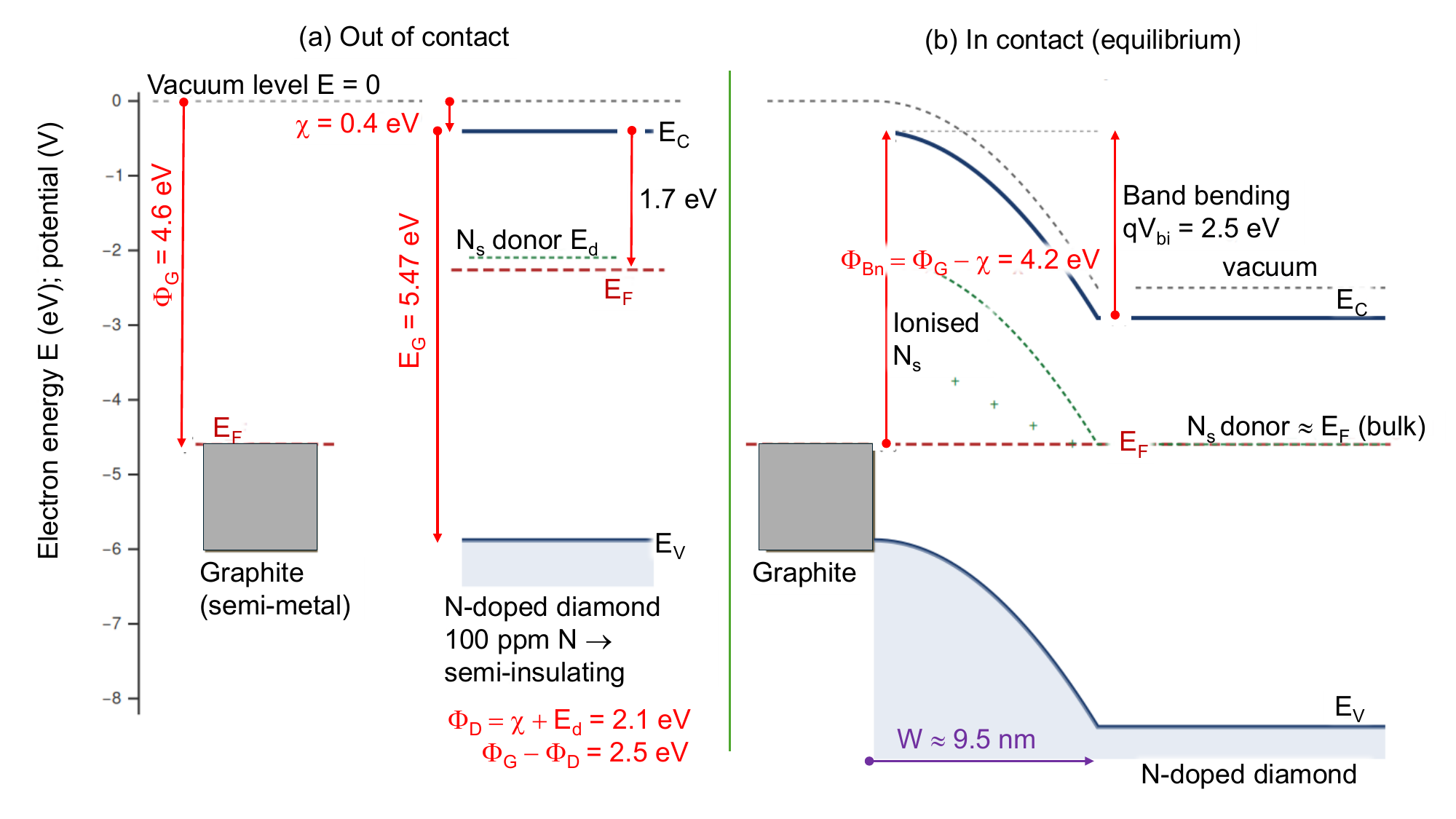%f-junk.pdf
} %{f-junk}
\caption{\label{f-junk} Energy band diagram out of (a) and in (b) contact between graphite and diamond doped by nitrogen at \SI{100}{ppm}. Standard energy notations for the electron work function, vacuum affinity, and bandgap edges are shown in the diagrams.  }
\end{figure*}

The boundary between laser-written graphite and the surrounding nitrogen-doped diamond
(Fig.~\ref{f-junk}) is a metal--semiconductor contact rather than a semiconductor heterojunction,
because graphite is a semimetal with no bandgap and a work function
$\Phi_G \approx \SI{4.6}{eV}$~\cite{ristein2000}. Within the Schottky--Mott picture the barrier seen
by electrons is $\Phi_{B,n} = \Phi_G - \chi$, where the electron affinity $\chi$ of diamond is set by
the surface dipole and ranges from $\chi \approx \SI{-1.3}{eV}$ for hydrogen termination to
$\chi \approx +\SIrange{1.3}{1.7}{eV}$ for oxygen termination~\cite{cui1998,riedel2004,rutter1998},
so the ideal barrier spans roughly \SIrange{2.9}{5.9}{eV}.

We do not convert this into a depletion width. Doing so requires the ionised donor density $N_D^+$
from a charge-neutrality solution including compensation, not the total nitrogen concentration $N_s$:
for a donor \SI{1.7}{eV} deep the great majority of centres are neutral at room temperature, and the
Ni-related compensation described in Sec.~\ref{s-exp} is unquantified, so an estimate using $N_s$ in
place of $N_D^+$ would overestimate the space charge and underestimate the depletion width. Nor are
vacuum-referenced affinities applicable without qualification to a buried, defect-dense
graphite/diamond interface at which the Fermi level is expected to be
pinned~\cite{ristein2000,zaitsev2001}, and no current--voltage data were taken on these structures.

The polarisation-loss analysis of Sec.~\ref{s-S} does not depend on any of this. An isolated
conductive inclusion dissipates energy in an oscillating field through its own internal conduction
current and requires no ohmic connection to the host; the junction would matter only for a percolation
or contact-resistance model, which is not advanced here.

%____________________________fig A1

\section{Temperature and spectrometer calibration}
\label{a-calib}

\begin{figure*}[h!]
\centering\includegraphics[width=.85\textwidth]{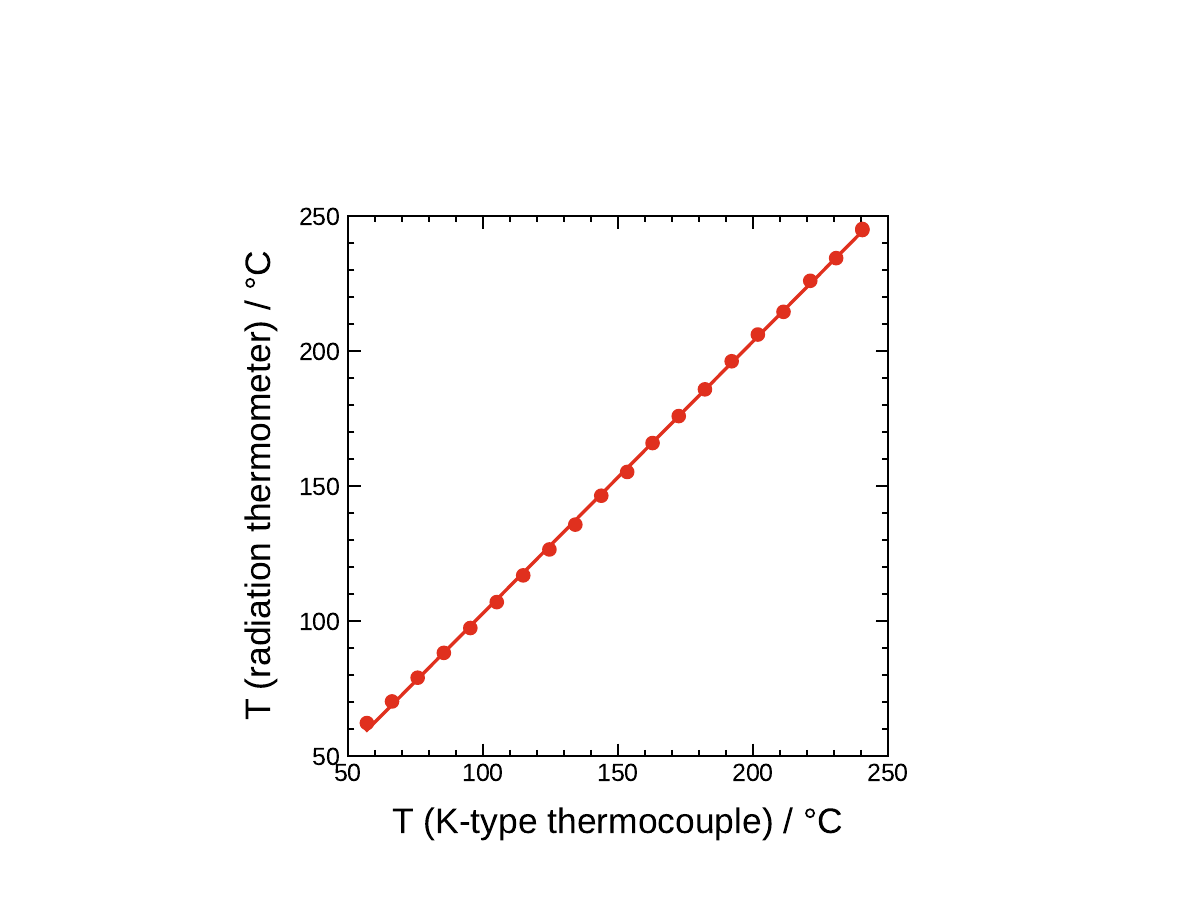}
\caption{\label{f-tk} Calibration of the radiative pyrometer using a standard type-K
Chromel--Alumel thermocouple on a silicon sample~\cite{13oe6901,16semsc221}. %The inset shows scanning electron microscopy (SEM) image of the side-view cross section of the black-Si sample with $\sim 290$~nm tall needles; orientation in the image corresponds to the ``vertical needles'' in the heating experiment.
}
\end{figure*}

Figure~\ref{f-tk} shows calibration of the radiative pyrometer against a standard type-K
Chromel--Alumel thermocouple, for flat Si. The microwave cavity was heated completely and uniformly using four PID-controlled cartridge heaters, with the setpoint increased in $10^\circ$C steps at 15-minute intervals after steady state has been reached according to both the thermocouple attached to the sample and the pyrometer. The linearity of the fit was confirmed with $R^2 > 0.99$ to $250^\circ$C. %The experiment also confirmed that the emissivity of the nano-smooth Si surface and of the nano-structured b-Si surface are the same.

\begin{figure*}[h!]
\centering\includegraphics[width=.8\textwidth]{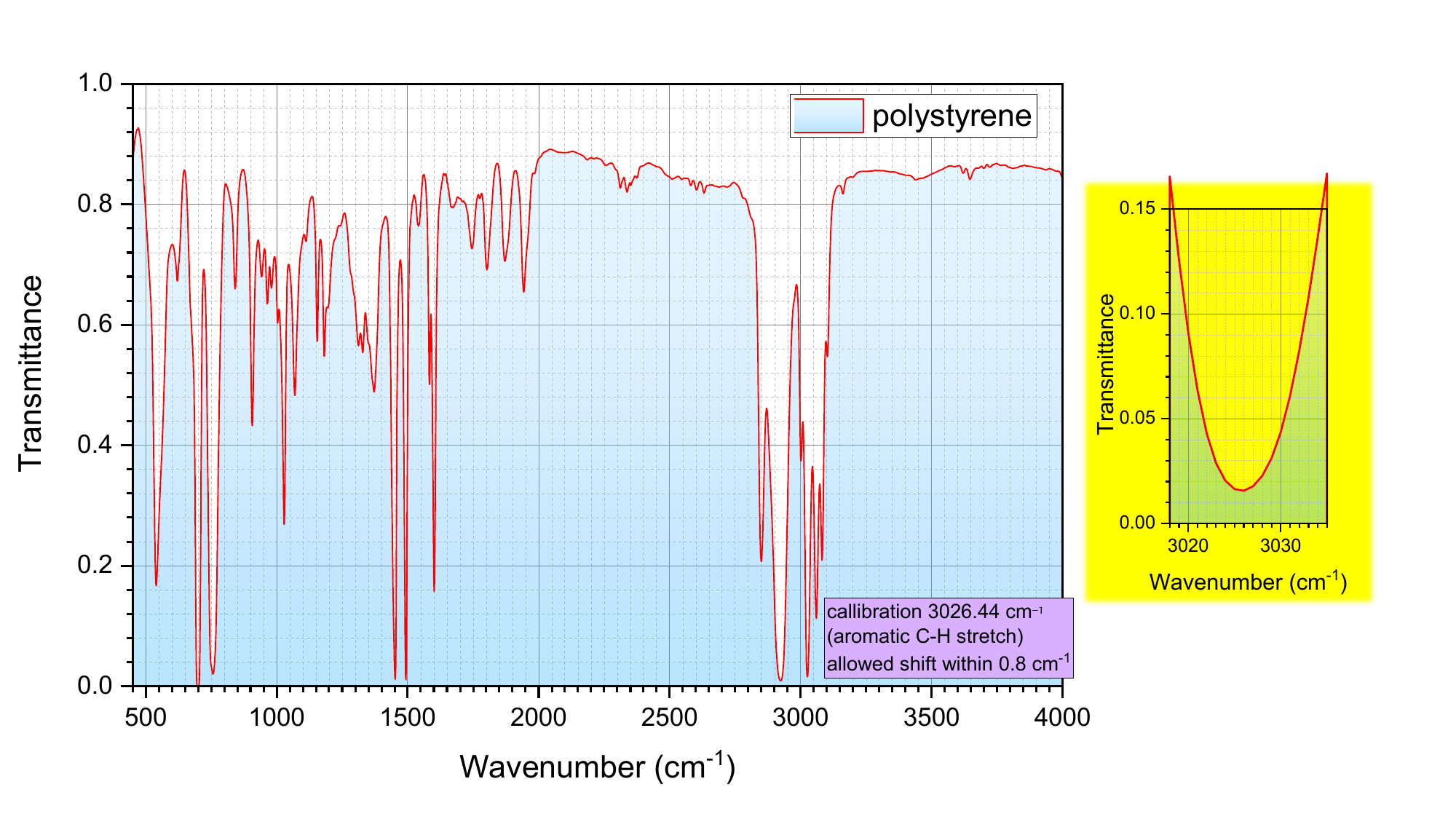}
\caption{\label{f-cali} Calibration of the IR spectrometer before the measurements shown in
Fig.~\ref{f-abs}(b).}
\end{figure*}

%The scope of this calibration should be stated, because the heating run of Sec.~\ref{s-results} is interpreted at indicated temperatures up to roughly four times the highest calibration point, on a different material whose optical state changes during the measurement. The calibration establishes that the instrument responds linearly against a thermocouple reference for silicon targets in this optical arrangement, up to $250^\circ$C. It establishes nothing about the spectral emissivity of diamond, of buried graphitic tracks viewed through diamond, of oxidised or roughened diamond surfaces, or of the quartz wool and cavity surfaces that share the \SI{3}{mm} measurement spot. ; and the finding that smooth and nanostructured silicon share an emissivity does not transfer to a diamond-to-graphite transition, which changes the optical constants rather than the surface morphology alone.%
Appendix~\ref{a-emis} quantifies the resulting exposure. Figure~\ref{f-cali} shows the calibration of
the IR spectrometer used for Fig.~\ref{f-abs}(b).

%____________________________fig A2

%____________________________fig A3

\section{Sensitivity of indicated temperature to emissivity}
\label{a-emis}

A single-band pyrometer infers temperature from spectral radiance assuming a fixed emissivity, here
effectively unity. If the true spectral emissivity is $\epsilon < 1$, the indicated temperature $T_a$
and the true temperature $T$ are related, in the Wien approximation, by
\begin{equation}
\frac{1}{T_a} = \frac{1}{T} - \frac{\lambda}{c_2}\ln\epsilon ,
\label{e-wien}
\end{equation}
with $c_2 = \SI{14388}{\micro\meter.K}$ the second radiation constant and $\lambda$ the effective
wavelength, taken as \SI{2.2}{\micro\meter} for the 1.95--2.6~$\mu$m band. Since $\ln\epsilon < 0$,
$T_a < T$ always: a low-emissivity target reads cold. Table~\ref{t-emis} evaluates
Eq.~(\ref{e-wien}) for a true surface temperature of $900^\circ$C.

\begin{table}[h!]
\caption{\label{t-emis} Indicated temperature for a true surface temperature of $900^\circ$C at
$\lambda = \SI{2.2}{\micro\meter}$, as a function of effective spectral emissivity.}
\begin{ruledtabular}
\begin{tabular}{cc}
Effective emissivity $\epsilon$ & Indicated temperature ($^\circ$C) \\
\colrule
0.05 & 490 \\
0.10 & 557 \\
0.30 & 692 \\
0.50 & 770 \\
0.80 & 855 \\
1.00 & 900 \\
\end{tabular}
\end{ruledtabular}
\end{table}

The span from $\epsilon = 0.1$ to $\epsilon = 0.8$ covers roughly $300^\circ$C of indicated
temperature at fixed true temperature, and precisely such an excursion is expected as the sample
passes from a transparent, low-emissivity yellow crystal to a roughened, graphitised and partially
oxidised surface. Reading Eq.~(\ref{e-wien}) in the other direction is more directly relevant: an
indicated $500^\circ$C from a surface of emissivity 0.1 corresponds to a true temperature of about
$790^\circ$C, and an indicated $900^\circ$C from a surface of emissivity 0.8 to about $950^\circ$C.
The observed excursion from an indicated 500 to an indicated $900^\circ$C is therefore consistent with
a true excursion of only some \SI{160}{K} if the emissivity rose from 0.1 to 0.8 over the same
interval. Applying the same relation to the detection floor, an indicated $40^\circ$C from a
low-emissivity surface corresponds to a true temperature near $79^\circ$C.

None of this makes the transition an artefact; the point is narrower and unavoidable. With a
single-band instrument and an optically evolving target, a temperature change and an emissivity change
enter the measurement identically and cannot be separated. Two-colour pyrometry addresses this
directly by forming a radiance ratio at two wavelengths, which is independent of emissivity for a grey
body; the grey-body assumption should itself be tested against a multispectral measurement.

\section{Q-factor change}

%____________________________fig A4
\begin{figure*}[b!]
\centering\includegraphics[width=.9\textwidth]{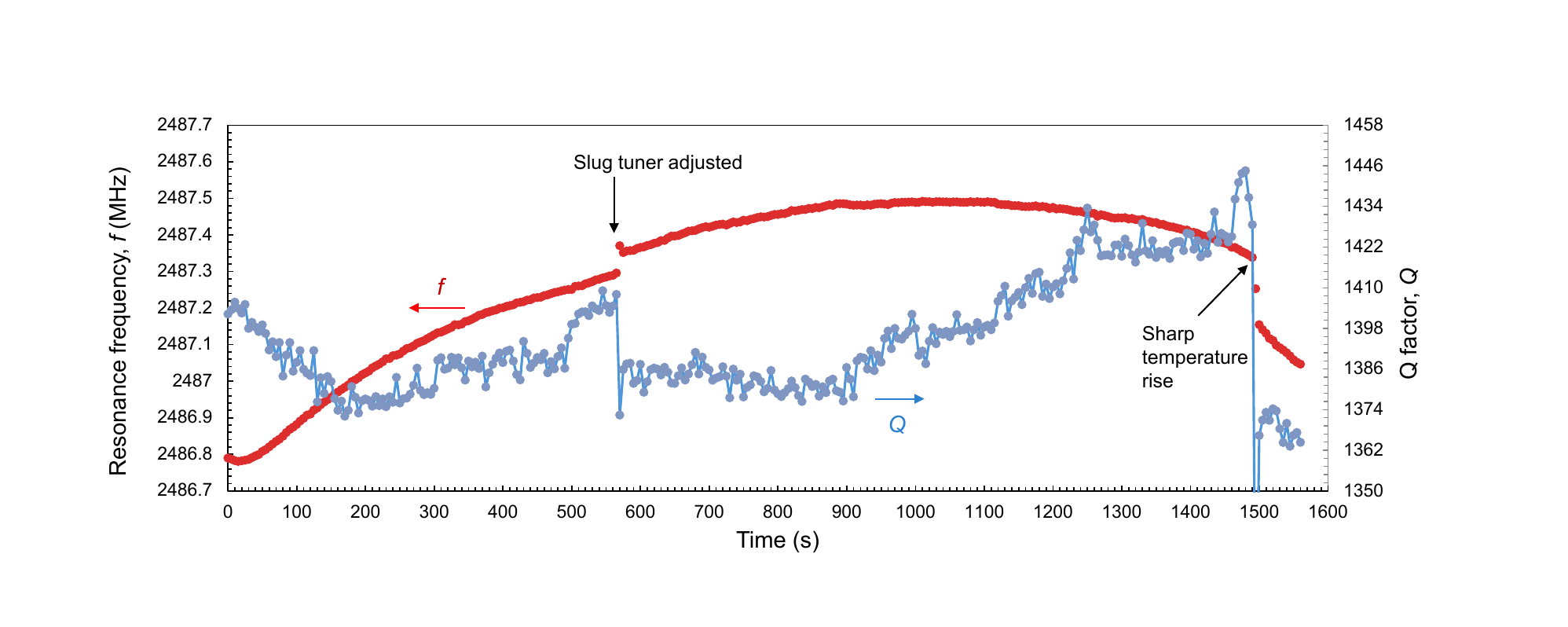}
\caption{\label{f-Q} Temporal evolution of the resonance frequency $f$ and quality factor $Q$ during the heating experiment. A pronounced decrease in both f and Q is observed following the final increase in microwave power, suggesting a temperature-induced transition in the electromagnetic coupling between the sample and the cavity. }
\end{figure*}

Time evolution of central resonance frequency and its $Q$-factor is plotted in Fig.~\ref{f-Q}. The sharp temperature rise at the end of heating and oxidation/burning initiation is linked to the sudden decrease of the $Q$-factor (increase of the width of the resonance).

\end{document}